\pdfoutput=1
\documentclass[cernpreprint,UKenglish,texlive=2016,texmf]{atlasdoc}
 
\usepackage[abbrevref=false]{atlaspackage}
\usepackage{atlasbiblatex}
 
\usepackage[unit=false, hepparticle=true]{atlasphysics}
\usepackage{multirow}
 
\graphicspath{{logos/}{fig/}}
 
\usepackage{ANA-TOPQ-2016-06-PAPER-defs}

\AtlasTitle{Measurement of single top-quark production in association with a $W$ boson in the single-lepton channel at  $\sqrt{s} = 8\,\text{TeV}$ with the ATLAS detector}
 
\author{The ATLAS Collaboration}
 
\AtlasAbstract{
The production cross-section of a top quark in association with a $W$ boson is measured
using proton--proton collisions at $\sqrt{s} = 8\,\text{TeV}$. The dataset
corresponds to an integrated luminosity of $20.2\,\text{fb}^{-1}$, and was collected
in 2012 by the ATLAS detector at the Large Hadron Collider at CERN.
The analysis is performed in the single-lepton channel. Events are selected by requiring one isolated
lepton (electron or muon) and at least three jets. A neural network is trained
to separate the $tW$ signal from the dominant $t\bar{t}$ background.
The cross-section is extracted from a binned profile
maximum-likelihood fit to a two-dimensional discriminant built from
the neural-network output and the invariant mass of the hadronically decaying $W$ boson.
The measured cross-section is $\sigma_{tW} = 26 \pm 7\,\text{pb}$,
in good agreement with the Standard Model expectation.
}

\AtlasRefCode{ANA-TOPQ-2016-06}

\PreprintIdNumber{CERN-EP-2019-221}

\arXivId{2007.01554}

\AtlasJournalRef{Eur.\ Phys.\ J.\ C 81 (2021) 720}
\AtlasDOI{10.1140/epjc/s10052-021-09371-7}

\hypersetup{pdftitle={ATLAS document},pdfauthor={The ATLAS Collaboration}}

\begin{document}
 
\maketitle
 
\tableofcontents
 
\newpage
 
 
\section{Introduction}
\label{sec:intro}
 
\begin{figure}[htbp]
\centering
\subfloat[$t$-channel]{\scalebox{0.5}{\includegraphics[scale=0.45]{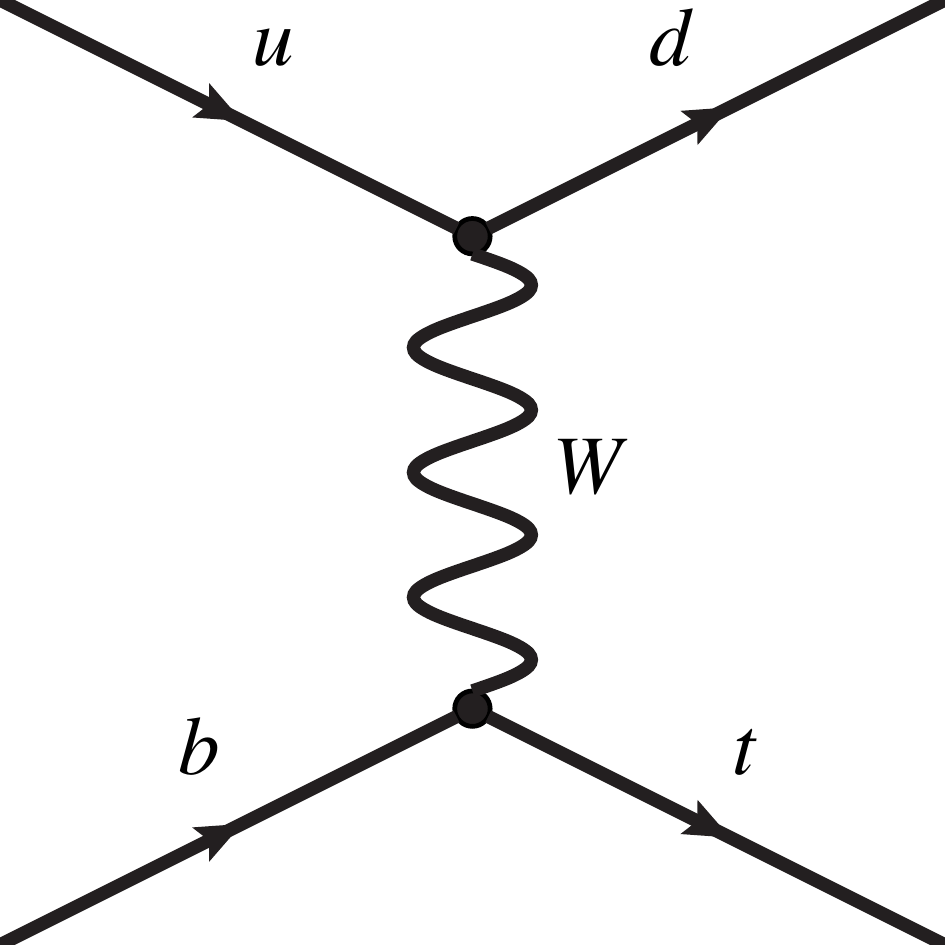}}}
\qquad
\subfloat[$s$-channel]{\scalebox{0.5}{\includegraphics[scale=0.40]{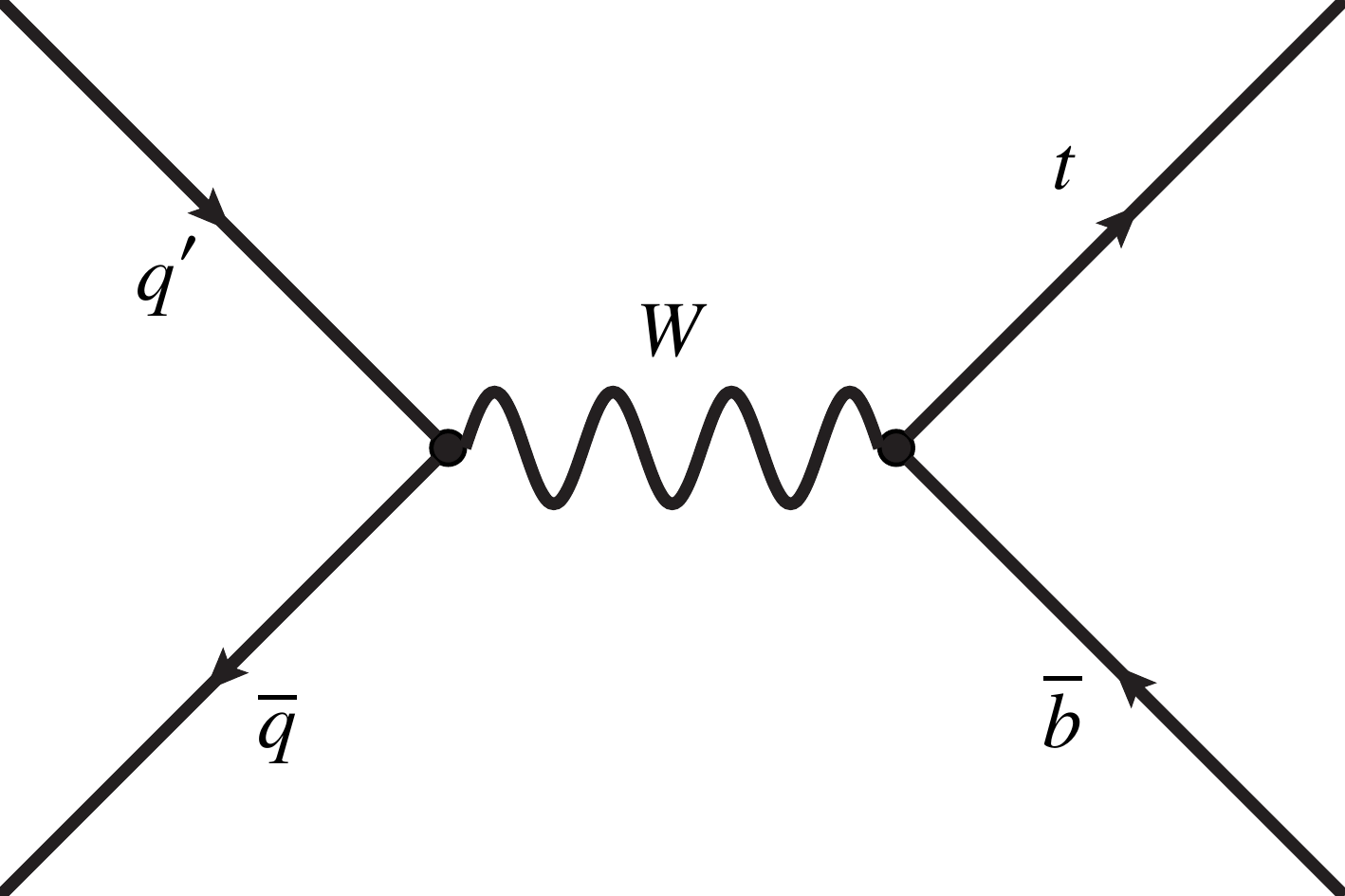}}}
\qquad
\subfloat[\Wt production]{\scalebox{0.5}{\includegraphics[scale=0.40]{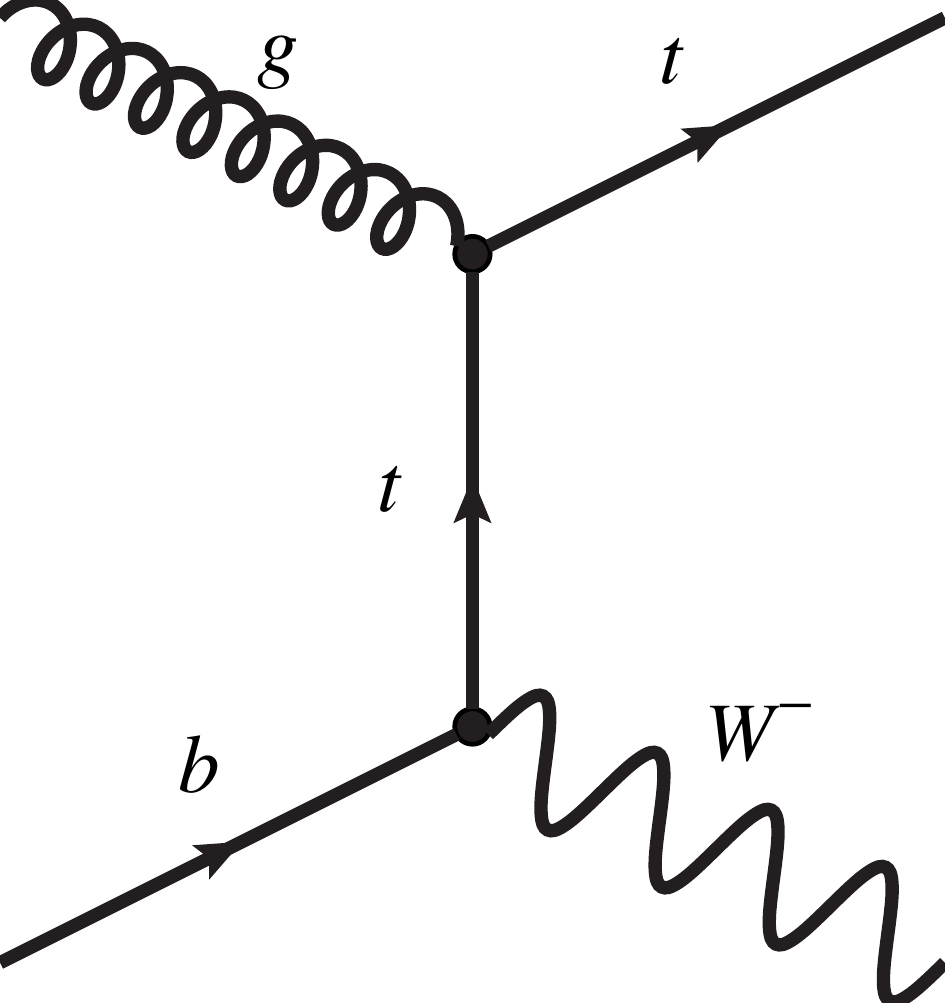}}}
\caption{Example LO Feynman diagrams of single top-quark
production: (a) $t$-channel,
(b) $s$-channel
and (c) \Wt production.}
\label{fig:SingleTopFeynmanDiagrams}
\end{figure}

Single top quarks are produced in proton--proton collisions via the weak interaction.
At leading order (LO) three different channels, which depend on
the virtuality of the \PW boson involved, are defined:
$t$-channel, $s$-channel or top-quark production in association with a \PW boson,
called \Wt production.
These processes, for which example Feynman diagrams are shown in \cref{fig:SingleTopFeynmanDiagrams}, involve a $Wtb$ vertex at LO in the Standard Model (SM).
Calculations involving \Wt production beyond LO have to include quantum interference with \ttbar production.
Measurements of single-top-quark cross-sections are used
to study the properties of this vertex,
as they are directly sensitive to the Cabibbo--Kobayashi--Maskawa
(CKM) matrix element \Vtb.
Deviations from the cross-sections predicted by the SM
can originate from single top quarks produced with similar kinematics
in the decays of
unknown heavy particles predicted by physics beyond the Standard Model.
If the masses of these particles are beyond the reach of direct searches, they might be revealed through their effects on the effective \Wtb{} coupling~\cite{Tait:2000sh}.
Using measurements in all three channels of single top-quark
production, physics beyond the SM can be probed
systematically in the context of Effective Field
Theory~\cite{Cao:2007ea}.
As each of the single-top-quark processes can be sensitive to different
sources of new physics, it is also important to study each
channel separately.
In addition, the SM production of \Wt is an important background in direct searches for
particles beyond the SM~\cite{SUSY-2016-29,SUSY-2016-16}.
 
At the Large Hadron Collider (LHC),
evidence for the \Wt production process was found by the ATLAS~\cite{TOPQ-2011-17}
and CMS Collaborations~\cite{CMS-TOP-11-022} at $\sqs =\SI{7}{\TeV}$
and the process was observed by both
experiments~\cite{TOPQ-2012-20,CMS-TOP-12-040} at $\sqs =\SI{8}{\TeV}$.
The \Wt cross-section has been also measured with \SI{13}{\TeV}
collision data inclusively by the CMS Collaboration~\cite{CMS-TOP-17-018}
as well as inclusively and differentially by the ATLAS Collaboration~\cite{TOPQ-2015-16,TOPQ-2016-12,TOPQ-2017-05}.
These measurements were performed in final states with two leptons, and the measured cross-sections
agree with the theoretical expectations.
 
This paper presents evidence for \Wt production in final states with a single lepton
using proton--proton (\pp) collisions at \sqs=~\SI{8}{\TeV}.
This topology features a \PW boson in addition to a top quark, which
decays mainly into another \PW boson and $b$-quark, leading to a $W^{+}W^{-}b$ state.
In the single-lepton channel, one of the \PW bosons decays
leptonically (\Wlep) while the other one decays
hadronically (\Whad). Therefore, the experimental signature of event
candidates is characterised by one isolated charged lepton (electron
or muon), large missing transverse momentum (\met), and three jets
with high transverse momentum (\pt), one of which contains a \Pqb-hadron and is
labelled as a \Pqb-tagged jet, \bj.
In contrast to the dilepton analyses, the event signature contains
only one neutrino, which originates from the leptonic \PW-boson decay.
Hence, both the \PW-boson and the top-quark kinematics can be
reconstructed and used to separate the signal from background.
The main backgrounds are \Wjets and \ttbar events, where the latter background
poses a major challenge in this measurement because of its similar
kinematics and a ten times larger cross-section compared to the \Wt signal.
An artificial neural network is trained to separate the signal from the
\ttbar background.
The cross-section is extracted using a binned profile maximum-likelihood fit to a two-dimensional discriminant.
This measurement, performed with \Wt single-lepton events, constitutes a cross-check of the previous results
published in the dilepton channel.
 
 
\newcommand{\AtlasCoordFootnote}{
ATLAS uses a right-handed coordinate system with its origin at the nominal interaction point (IP)
in the centre of the detector and the $z$-axis along the beam pipe.
The $x$-axis points from the IP to the centre of the LHC ring,
and the $y$-axis points upwards.
Cylindrical coordinates $(r,\phi)$ are used in the transverse plane,
$\phi$ being the azimuthal angle around the $z$-axis.
The pseudorapidity is defined in terms of the polar angle $\theta$ as
$\eta = -\ln \tan(\theta/2)$.
Angular distance is measured in units of $\Delta R \equiv \sqrt{(\Delta\eta)^{2} + (\Delta\phi)^{2}}$.}
\section{ATLAS detector}
\label{sec:atlas}
 
The ATLAS experiment~\cite{PERF-2007-01} at the LHC is a multipurpose particle detector
with a forward--backward symmetric cylindrical geometry and a near $4\pi$ coverage in
solid angle.\footnote{\AtlasCoordFootnote}
It consists of an inner tracking detector (ID) surrounded by a thin superconducting solenoid
providing a \SI{2}{\tesla} axial magnetic field, electromagnetic and
hadron calorimeters, and a muon spectrometer (MS).
The ID provides charged-particle tracking in the pseudorapidity range $\abseta < 2.5$.
It consists of silicon pixel, silicon microstrip, and transition-radiation tracking detectors.
Lead/liquid-argon (LAr) sampling calorimeters provide electromagnetic (EM) energy measurements
with high granularity.
An iron/scintillator-tile hadron calorimeter covers the central pseudorapidity range ($|\eta| < 1.7$).
The endcap  ($1.5<\abseta <3.2$) and forward  ($3.1<\abseta<4.9$) regions are instrumented with LAr calorimeters
for measurements of both EM and hadronic energy.
The MS surrounds the calorimeters and includes a system of precision tracking chambers ($|\eta| < 2.7$) and fast detectors for triggering ($|\eta| < 2.4$).
The magnet system for the MS consists of three
large air-core toroidal magnets with eight superconducting coils.
The field integral of the toroids ranges between \num{2.0} and \SI{6.0}{\tesla\metre}
across most of the detector.
Collisions producing interesting events are selected for storage with the trigger system~\cite{PERF-2011-02}.
For the data taken at $\sqrt{s} = \SI{8}{\TeV}$,
a three-level trigger system was used to select events.
The first-level trigger is implemented in hardware and uses a subset of the detector information.
It reduced the accepted rate to at most \SI{75}{\kilo\hertz}.
This was followed by two software-based trigger levels that
together reduced the accepted event rate to \SI{400}{\hertz} on average,
depending on the data-taking conditions.
 
 
\section{Data and simulated event samples}
\label{sec:dataMC}
The data considered in this analysis are from
\pp collisions at  $\sqs=\SI{8}{\TeV}$ and were
taken with stable LHC beams and the ATLAS
detector fully operational, corresponding to
an integrated luminosity of \SI{20.2}{\per\femto\barn}.
 
Monte Carlo (MC) samples were produced using the full ATLAS
detector simulation~\cite{SOFT-2010-01}  implemented in \GEANT~\cite{Agostinelli:2002hh}.
In addition, alternative MC samples, used to train the neural network and
evaluate systematic uncertainties, were produced using
\AtlFast~\cite{SOFT-2010-01},
which provides a faster calorimeter simulation making use of parameterised
showers to compute the energy deposited by the particles.
Pile-up (additional \pp interactions in the same or nearby bunch
crossing)
was modelled by overlaying simulated minimum-bias
events generated with  \PYTHIA[8]~\cite{Sjostrand:1064095}.
Weights were assigned to the simulated events,
such that the distribution of the number of \pp interactions per bunch crossing in
the simulation matches the corresponding
distribution in the data, which has an average of 21~\cite{DAPR-2013-01}.
 
The \Wt signal events were simulated using the next-to-leading order (NLO) \POWHEG method~\cite{1126-6708-2004-11-040, 1126-6708-2007-11-070, Re:2010bp}
implemented in the \POWHEGBOX[(v.1.0)] generator  (revision 2192)~\cite{Alioli2010}
with the CT10 parton distribution function (PDF) set~\cite{PhysRevD.82.074024} in the matrix-element calculation.
The mass and width of the top-quark were set to $m_{t} = \SI{172.5}{\GeV}$ and $\Gamma_{t} = \SI{1.32}{\GeV}$, respectively.
The top quark was assumed to decay exclusively into $\PW\Pqb$.
The parton shower, hadronisation and underlying event were simulated
using \PYTHIA[6] (v6.426)~\cite{1126-6708-2006-05-026} with the LO
CTEQ6L1 PDF set~\cite{Nadolsky:2008zw} and a corresponding set of tuned parameters
called the Perugia 2011 (P2011C) tune~\cite{Skands:2010ak}. The
factorisation scale, \muF, and renormalisation scale, \muR, were set to $m_t$.
Calculations involving \Wt production beyond LO included quantum
interference with \ttbar production.
Double-counting of the contributions was avoided by using either the diagram-removal (DR) or the diagram-subtraction (DS) scheme~\cite{Frixione:2008yi,White:2009yt}.
In the DR scheme, diagrams with a second on-shell top-quark propagator are removed from the amplitude,
while in the DS scheme, a subtraction term cancels out the \ttbar contribution to the cross-section when the top-quark propagator becomes on shell.
Nominal MC samples were generated using the DR scheme.
For the evaluation of systematic uncertainties, alternative samples were generated using the DS scheme,
or using \POWHEGBOX or \MCatNLO[(v4.06)]~\cite{1126-6708-2002-06-029}, each interfaced with \HERWIG[(v6.520)]~\cite{1126-6708-2001-01-010}.
For the \HERWIG samples, the AUET2 tune~\cite{ATL-PHYS-PUB-2011-008} with the CT10 PDF was used
and the underlying event was generated with \JIMMY[(v4.31)]~\cite{Butterworth1996}.
In addition, \PYTHIA[6] (v6.427) samples with variations of \muR and \muF and the radiation tunes were used.
The SM \Wt cross-section prediction at NLO including next-to-next-to-leading-log~(NNLL) soft gluon corrections~\cite{Kidonakis:2010ux, Kidonakis:2013zqa} was calculated as
$\sigma_{\Wt}^{\text{th.}}(\SI{8}{\TeV})=22.4 \pm 0.6\,(\text{scale})\pm{1.4}\,(\text{PDF})~\si{\pb}$
assuming a top-quark mass, $m_t$, of $\SI{172.5}{\GeV}$.
The first uncertainty accounts for renormalisation and
factorisation scale variations (from $m_t/2$ to $2m_t$)
and the second term covers the uncertainty in the parton distribution functions,
evaluated using the MSTW2008 PDF set~\cite{Martin2009} at next-to-next-to-leading order (NNLO).
 
The \ttbar sample was generated with \POWHEGBOX[(v1.1)] interfaced
with \PYTHIA[6] (v6.427)~\cite{Frixione:2007nw}.
In the \POWHEGBOX event generator, the CT10 PDFs
were used, while the CTEQ6L1 PDFs were used for \PYTHIA{}.
The $h_{\text{damp}}$ parameter, which effectively regulates the high-\pt
gluon radiation, was set to $m_t$.
The predicted \ttbar production cross-section,
$\sigma_{\Pqt\Paqt} (\SI{8}{\TeV})=252.9^{+6.4}_{-8.6}\,\text{(scale)} \pm 11.7\,(\text{PDF+\alphas})\,\si{\pico\barn}$,
was calculated with the Top++2.0 program to NNLO in perturbative QCD,
including soft-gluon resummation to NNLL~\cite{Czakon:2011xx}.
The first uncertainty comes from the sum in quadrature of the effects of independently
varying \muR  and \muF. The uncertainty associated with
variations in the PDFs and strong coupling constant, \alphaS, was evaluated following the
PDF4LHC NLO prescription~\cite{Botje:1318947, Demartin:2010er}, which defines the central value
as the midpoint of the uncertainty envelope of three PDF sets:
MSTW2008 NNLO~\cite{Martin2009}, CT10 NNLO~\cite{PhysRevD.89.033009}
and NNPDF2.3~5f~FFN~\cite{Ball:2012cx}.
The same procedures as for the \Wt samples were employed to determine
the uncertainties due to the NLO matching method and the parton shower
and hadronisation.  Samples to evaluate the scale uncertainties were
produced in a similar way, varying \muR and \muF together with the
\Perugia tune, but also adding variations in the $h_{\text{damp}}$ parameter (for the up-variation, $h_{\text{damp}}$ was changed to $2 m_t$,
while for the down variation it was kept at $m_t$).
 
The other single-top-quark production processes, $s$-channel and
$t$-channel, were also generated with \POWHEGBOX[(v1.1)] coupled to
\PYTHIA[6] (v6.426), using the same PDF sets as described for the other top-quark
processes above.
The predicted cross-sections at \sqs=\SI{8}{\TeV} were calculated at NLO plus NNLL as
$5.6\pm\SI{0.2}{\pico\barn}$
for the $s$-channel~\cite{Kidonakis:2010tc, Alioli:2009je},
and $87.8^{+3.4}_{-1.9}\,\si{\pb}$ for the $t$-channel~\cite{Kidonakis:2011wy, Frederix:2012dh} process.
 
The multi-leg LO generator \SHERPA[(v1.4.1)]~\cite{Gleisberg:2008ta,Hoeche:2009rj,Schumann:2007mg},
together with the CT10 PDF sets,
was used to simulate vector-boson production in association with jets.
\SHERPA was used  to generate the hard process as
well as the parton shower and the modelling of the underlying event.
Double-counting between the inclusive $V+n$ parton samples (with $V=$ \PW or \PZ) and samples with associated
heavy-quark pair production was avoided consistently by using massive \Pqc- and \Pqb-quarks in the shower.
The predicted NNLO \Wjets cross-section with \PW decaying leptonically
was calculated as
$\xs(\Pp{}\Pp \rightarrow\Plpm\Pgnl X) = 36.3\pm\SI{1.9}{\nb}$~\cite{PhysRevD.69.094008}.\footnote{
The quoted cross-section is the sum of decays into \Pe, \Pgm and \Pgt.} 
For \Zjets the cross-section was calculated at NNLO in QCD for leptonic \PZ
decays as $\xs(\pp \rightarrow \Plp{}\Plm X) = 3.72\pm\SI{0.19}{\nb}$~\cite{PhysRevD.69.094008}. 
The \AtlFast simulation was used to generate these samples with sufficient statistics.
For cross-checks of the \Wjets{} modelling, an alternative sample generated with \ALPGEN[(v2.14)]~\cite{Mangano:2002ea} with up to five additional partons, \PYTHIA[6] (v6.426) and the CTEQ6L1 PDFs were used.
Diboson samples (\VVjets) were generated with \HERWIG[(v6.520)] at LO QCD using the CTEQ6L1 PDF.
The theoretical NLO cross-section for events with one lepton
is $29.4\pm\SI{1.5}{\pico\barn}$~\cite{Campbell:1999ah}.

Multijet events are selected in the analysis when they contain
jets or photons misidentified as leptons
or contain non-prompt leptons from hadron decays  (both referred to as a
\enquote{fake} lepton).
This background  was estimated directly from data using the matrix
method~\cite{ATLAS-CONF-2014-058}, which exploits differences in lepton
identification and isolation properties between prompt and fake leptons.
The data were processed with a second, \enquote{loose} set of lepton selection criteria.
The resulting sample was then corrected for efficiency differences between the two sets of cuts, and the contamination from events containing prompt leptons was subtracted.
The efficiencies, lepton selection criteria, and uncertainties applied in this analysis are the same as in Ref.~\cite{ATLAS-CONF-2014-058}.
 
 
\section{Object definitions}
\label{sec:objects}
 
Primary vertex (PV) candidates in the interaction region are
reconstructed from at least five tracks that satisfy a transverse
momentum ($\pt$) of $\pt>\SI{400}{\MeV}$.
The candidate with the highest sum of $\pt^{2}$ over all associated tracks is chosen as the hard-collision PV~\cite{PERF-2015-01}.
 
Muon candidates are reconstructed by matching segments or tracks in the
MS with tracks found in the ID~\cite{PERF-2014-05}.
The candidates must have $\pt > \SI{25}{\GeV}$ and be in the
pseudorapidity range $\abseta < 2.5$.
The longitudinal impact parameter of the track relative to the
hard-collision PV, $|\zvtx|$, is required to be smaller than \SI{2}{\mm}.
In order to reject non-prompt muons, an isolation criterion is applied.
The isolation variable is defined as  the scalar sum of the transverse
momenta of all tracks with $\pt > \SI{1}{\GeV}$ (excluding the muon
track) within a cone of size $\Delta R = \SI{10}{\GeV} / \pt(\mu)$ around the muon's direction.
It is required to be less than \SI{5}{\%} of the muon \pT. 
The selection efficiency after this requirement is measured to be about \SI{97}{\%} in $\PZ \to \Pgmp{}\Pgmm$ events.
 
Electron candidates are reconstructed from energy deposits (clusters) in the EM calorimeter,
which match a well-reconstructed track in the ID~\cite{PERF-2016-01}.
Requirements on the transverse and longitudinal impact parameter of
$|\dvtx | < \SI{1}{\mm}$ and $|\zvtx | < \SI{2}{\mm}$, respectively, are applied.
Electron candidates must have energy in the transverse plane $\et > \SI{25}{\GeV}$ and
$\abs{\eta_{\text{cluster}}} < 2.47$,  where $\eta_{\text{cluster}}$
denotes the pseudorapidity of the cluster.
Clusters in the calorimeter barrel--endcap
transition region,  $1.37 < \abseta < 1.52$, are excluded.
An isolation requirement based on the deposited transverse
energy in a cone of size $\Delta R = 0.2$ around the direction of the
electron and the \pt sum of the tracks in a
cone with $\Delta R = 0.3$ around the same direction is applied. This requirement
is chosen to give a nearly uniform selection efficiency of 85\% in \pt and $\eta$, as measured in $\PZ \to \Pep{}\Pem$ events.
Electron candidates that share the ID track with a
reconstructed muon candidate are vetoed.
 
Jets are reconstructed using the \antikt algorithm~\cite{Caciari:2008gp, Fastjet}
with a radius parameter of $R=0.4$
using topological clusters~\cite{PERF-2014-07}, calibrated with
the  Local Cluster Weighting method~\cite{PERF-2014-02},
as input to the jet finding.
The jet energy is further corrected by subtracting the contribution
from pile-up events and applying
an MC-based and a data-based calibration.
The jet vertex fraction (JVF) ~\cite{PERF-2014-03} variable is used to
identify the primary vertex from which the jet originated.
The JVF criterion suppresses pile-up jets with $\pt < \SI{50}{\GeV}$ and $\abseta < \num{2.4}$.
To avoid possible overlap between jets and electrons,
jets that are closer than $\Delta R = 0.2$ to an electron are removed.
Afterwards, remaining electron candidates overlapping with
jets within a distance of $\Delta R = 0.4$ are rejected.
Finally, muons overlapping with jets within $\Delta R = 0.4$ are removed.
 
The identification of jets originating from the hadronisation of a
\Pqb-quark (\Pqb-tagging) is based on  various algorithms exploiting
the long lifetime, high mass and high decay multiplicity of
\Pqb-hadrons 
as well as the properties of the \Pqb-quark fragmentation.
The outputs of these algorithms are combined in a neural network classifier to maximise the \Pqb-tagging performance~\cite{PERF-2012-04}.
The choice of \btag working point represents a trade-off between
the efficiency for identifying \Pqb-jets and rejection of other jets.
The chosen working point for this analysis corresponds to a \btag efficiency of
70\%. The corresponding \Pqc-quark-jet rejection factor
is about \num{5} and the light-quark-jet rejection factor is
about \num{120}.
These efficiencies and rejection factors were obtained using
simulated \ttbar events.
The tagging efficiencies in the simulation are corrected to match the efficiencies
measured in data~\cite{PERF-2012-04}.
 
The \(\pT^{\text{miss}}\) of the event, defined as the momentum imbalance
in the plane transverse to the beam axis, is primarily due to neutrinos that escape detection.
It is calculated as the negative vector sum
of the transverse momenta of the reconstructed electrons, muons, jets and the
clusters that are not associated with any of the previous objects
(the `soft term')~\cite{PERF-2014-04}.
Its magnitude is denoted \MET.
 
 
 
\section{Event selection}
\label{sec:selection}
 
Events are required to have a hard-collision primary vertex.
They also have to pass a single-lepton trigger requirement~\cite{PERF-2011-02,TRIG-2012-03}
and contain at least one electron or muon candidate with $\pT > \SI{30}{\GeV}$
matched to the lepton that fired the trigger.
The electron trigger requires an electron candidate, formed by an EM calorimeter cluster matched with a track, either with $\et > \SI{60}{\GeV}$
or with $\ET > \SI{24}{\GeV}$ and additional isolation requirements.
The muon trigger requires a muon candidate, defined as a reconstructed
track in the muon spectrometer, either with $\pT > \SI{36}{\GeV}$ or with
$\pT > \SI{24}{\GeV}$ and isolation requirements.
If there is another lepton candidate with a transverse momentum
above $\SI{25}{\GeV}$, the event is rejected.
This lepton veto guarantees
orthogonality with respect to the dilepton analysis.
The contribution from leptonically decaying \Pgt-leptons is included.
In the following, the electron or muon candidate is referred to as the lepton.
 
Events identified as containing jets from cosmic rays or beam-induced
backgrounds or due to noise hot spots in the calorimeter are removed.
Only jets with $\pT > \SI{30}{\GeV}$ and $\abseta < \num{2.4}$ are
considered in the analysis.
Additionally, a requirement of $\met > \SI{30}{\GeV}$ is applied, and the transverse
mass\footnote{
The transverse mass is calculated using the momentum of the lepton associated with the
\PW boson, \(\pT^{\text{miss}}\) and the azimuthal angle between the two:
$\mtwLep = \mtw =\sqrt{
2\pt(\Pl)\cdot\met\big[1-\cos(\Delta\phi(\Pl, \pT^{\text{miss}}))\big]
}$.
} of the leptonically decaying \PW boson must satisfy $\mtwLep >\SI{50}{\GeV}$.

In order to perform the measurement and validate the result, selected
events  are divided into different categories based on the jet and
$b$-tagged jet multiplicities.
The region with three jets of which one is $b$-tagged
(3j1b)
is called the signal region and is used to extract the \Wt cross-section.
The region with four jets, two of them
$b$-tagged (4j2b), contains a very pure sample of \ttbar
events and is used as the \ttbar validation region to check the modelling of this background.
Table~\ref{tab:prefit-yields_onlytW} shows the expected
and the observed
numbers of events in the signal region after the event selection.
All backgrounds except fake leptons, which is estimated using data-driven methods,
are normalised to their  expected cross-sections.
The \Wt events constitute about 5\% of the total number of events.
The major backgrounds are \ttbar production with about 58\%,
and \Wjets production with about 28\% of the total
number of events.
The \Wjets events are subdivided into heavy flavour (HF), where a \PW boson is produced in association with \Pqb- or \Pqc-jets,
and light flavour (LF).
The total numbers of expected events agree within a few percent with
the observed numbers of events.
 
\begin{table}[htbp]
\centering
\caption{Expected signal and background and observed number of
events in the signal 3j1b region. 
The cross-section for \Wt production is taken to be the theory prediction.
The uncertainties include statistical and systematic uncertainties.
}
\setlength{\tabcolsep}{3pt}
\begin{tabular}{
l
S[table-format=6.0,group-minimum-digits=0]@{$\,\pm\,$}
S[table-format=5.0,group-minimum-digits=0]
}
\toprule
\multicolumn{1}{c}{Process} & \multicolumn{2}{c}{Signal region} \\
\multicolumn{1}{c}{} & \multicolumn{2}{c}{(3j1b)} \\
\midrule
\Wt{} ($\sigma_{\Wt}^{\text{th.}}=\SI{22.4}{\pb}$) & 6300 & 600 \\
\midrule
\ttbar{} & 77000 & 6000 \\
$t$, $t$-channel & 4180 & 290\\
$t$, $s$-channel & 307 & 19 \\
\Wjets{}, HF & 31000 & 14000 \\
\Wjets{}, LF & 6000 & 3000  \\
\Zjets{} & 3900 & 1700  \\
\VVjets{} & 650 & 280 \\
Fake leptons & 4300 & 1900 \\
\midrule
Total background & 128000 & 18000 \\
\midrule
Total signal + background  & 134000 & 18000\\
Observed & \multicolumn{1}{S[table-format=6.0,group-minimum-digits=0]@{\phantom{$\,\pm\,$}}}{134633} & \\
\bottomrule
\end{tabular}
 
\label{tab:prefit-yields_onlytW}
\end{table}
 
 
 
\section{Separation of signal from background}
\label{sec:s_b_NN_discriminant}
 
Differences between signal and background event kinematics
are exploited to better separate them.
The \ttbar{} background is inherently difficult to distinguish from the signal,
motivating the use of an artificial neural network (NN) implemented in the NeuroBayes framework~\cite{Feindt:2006pm,Feindt:2004}.
Detailed information about how the NN is used in single-top-quark
analyses can be found in Ref.~\cite{TOPQ-2012-21}.
The NN input variables are selected such that they contribute significantly to the
statistical separation power between signal and background, while avoiding
variables that would lead to an increase of the expected uncertainty in the signal cross-section.
The observable \mjj{} (\Fig{\ref{fig:mjj_shape}}) provides very good separation of the signal from the background,
but is strongly affected by uncertainties in the reconstructed jet energies as well as uncertainties in the $b$-tagging in \ttbar{} events.
For this reason, \mjj{} is not used in the NN; instead a two-dimensional
discriminant is constructed from \mjj{} and the response of the NN.
The two-dimensional discriminant, explained in the following subsections, allows the
nuisance parameters affecting the variable \mjj to be partially constrained.

\subsection{Invariant mass of the hadronically decaying \PW boson}
\label{sec:had_mjj}
 
\begin{figure}[htbp]
\centering
\subfloat[]{\includegraphics[scale=0.5]{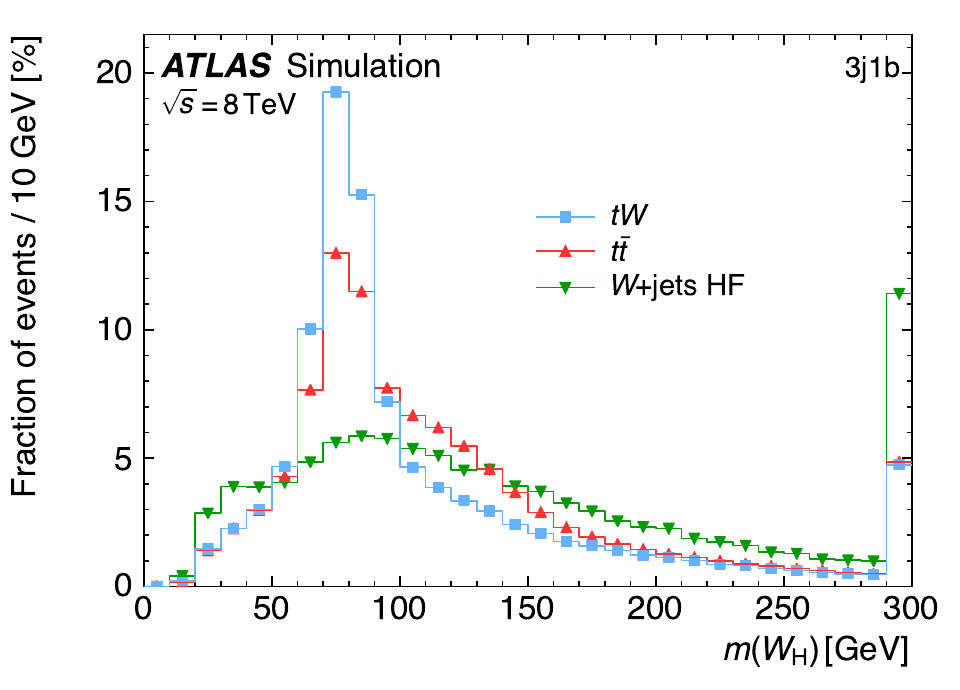}}
\subfloat[]{\includegraphics[scale=0.5]{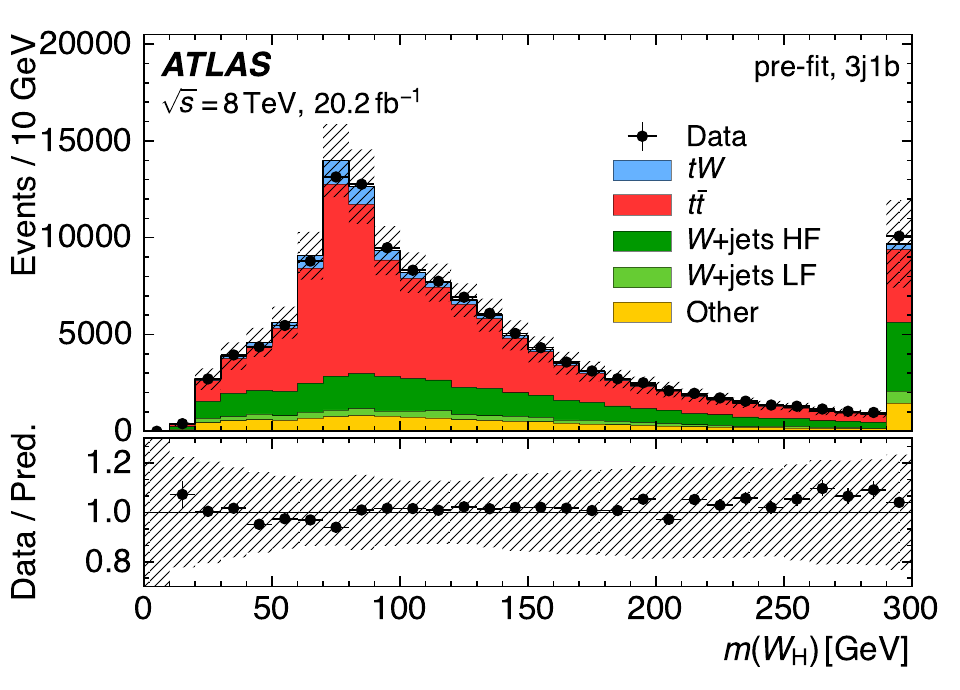}}
\caption{(a) Shape of the reconstructed \mjj distribution for
signal and most important backgrounds in the signal (3j1b) region.
\CaptionShape
(b) Pre-fit \mjj distribution in the 3j1b region.
\CaptionStack
\CaptionStackRatio
\CaptionStackOverflow
}
\label{fig:mjj_shape}
\end{figure}

The variable \mjj is computed from the four-momenta of the two
selected untagged jets.
For the signal and the \ttbar{} background,
the distribution of \mjj exhibits a peak near the mass of the \PW boson,
shown in \cref{fig:mjj_shape}(a).
The peak results from events where the two untagged jets are correctly
matched to the hadronically decaying \PW boson.
This is less likely to happen for \ttbar events than for \Wt events
due to the higher \Pqb-jet multiplicity and the limited \Pqb-tagging efficiency. On the other hand,
the \Wjets background does not feature such a peak since the \PW boson must
decay leptonically for the events to pass the selection.
\Cref{fig:mjj_shape}(b) shows the pre-fit distribution
of \mjj, and also demonstrates good pre-fit modelling of the data.
 
\subsection{Neural network}
\label{sec:neural_network}
The NN is trained using simulated events with the two reconstructed untagged
jets matched within $\Delta R < \num{0.35}$ to the generator-level jets originating
from a \PW-boson decay in the MC simulation and having a reconstructed mass of $\SI{65}{\GeV} < \mjj < \SI{92.5}{\GeV}$.
As events are required to contain a lepton, only \Wt, \ttbar{} and diboson events can have a pair of jets matched to the hadronic \PW-boson decay.
Given that the contribution from diboson production is very small, the background sample used for the training
consists entirely of \ttbar events.
Following the training procedure mentioned before,
the following four variables (ordered by significance) are selected as input for the NN:
 
\begin{itemize}
\item the transverse momentum of the \Wt system,
$\pt(\Whad\Wlep\bj)$, divided by the sum of the objects' transverse momenta,
\begin{linenomath}
\begin{equation*}
\Wtptbalance     = \frac{\pt(\Whad\Wlep\bj)}{\pt(\Whad)+\pt(\Wlep)+\pt(\bj)}\,,
\end{equation*}
\end{linenomath}
where the four-momentum of \Wlep is the sum of the four-momenta of the electron or muon and the neutrino, and
the four-momentum of the neutrino is determined using \met from the solution of a quadratic equation.\footnote{
There are two solutions if $\mtw < \mW$ and no real-valued solutions if $\mtw > \mW$.
In the first case, the ambiguity is resolved by picking the solution with the smaller $|p_z(\Pgn)|$.
The latter case occurs if the measured \met is too large,
and is resolved by adjusting \met until a real-valued solution is found,
under the constraint that the adjustment be minimal in the transverse plane.}
The use of \Wtptbalance, instead of the
transverse momentum of the \Wt system, decreases the background
contribution  in the signal-like region of the NN response and results in a gain
of sensitivity;
\item the invariant mass of the reconstructed \Wt system, $m\of\Wtsys$;
\item the absolute value of the difference between the pseudorapidities of the lepton and the
leading untagged jet in \pt, $|\Delta\eta(\ell,j_{L1})|$\,;
\item the absolute value of the pseudorapidity of the lepton, $|\eta(\ell)|$.
\end{itemize}
 
\begin{figure}[tbp]
\centering
\subfloat[]{\includegraphics[scale=0.5]{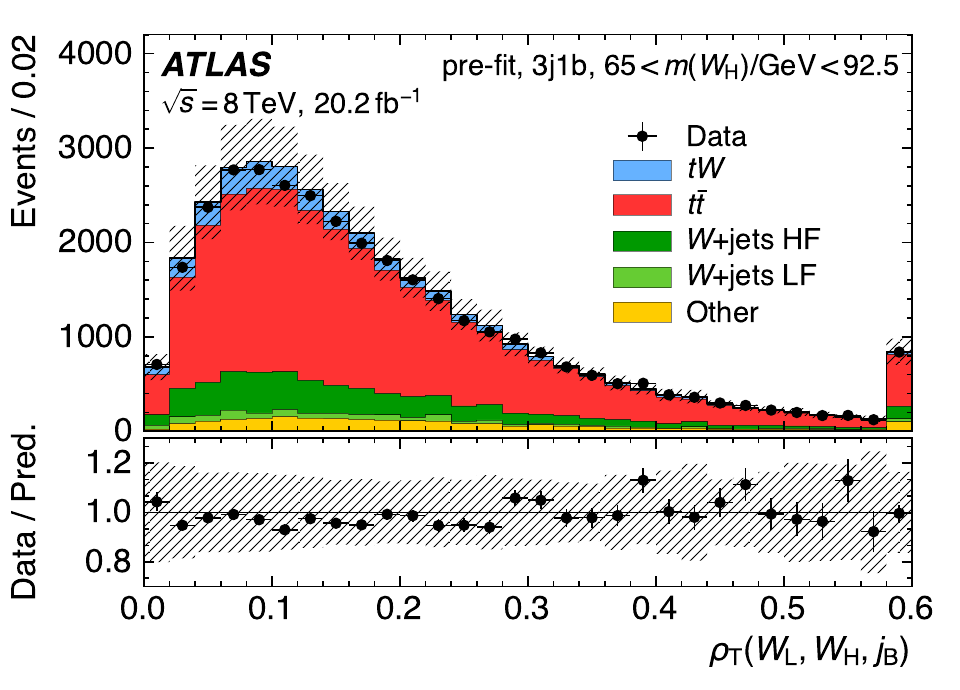}}
\subfloat[]{\includegraphics[scale=0.5]{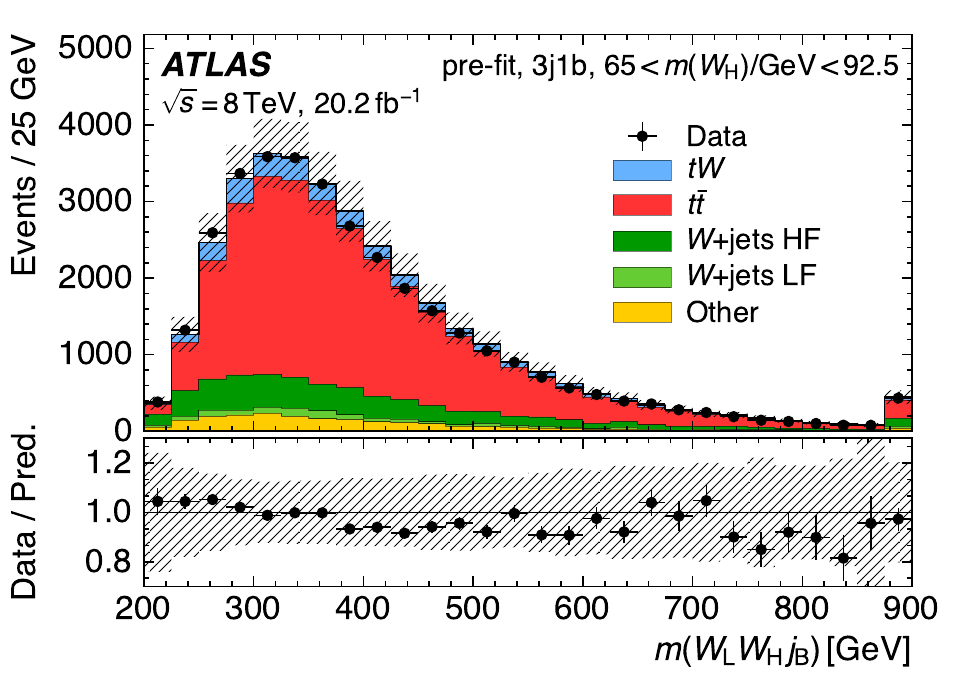}}\\
\subfloat[]{\includegraphics[scale=0.5]{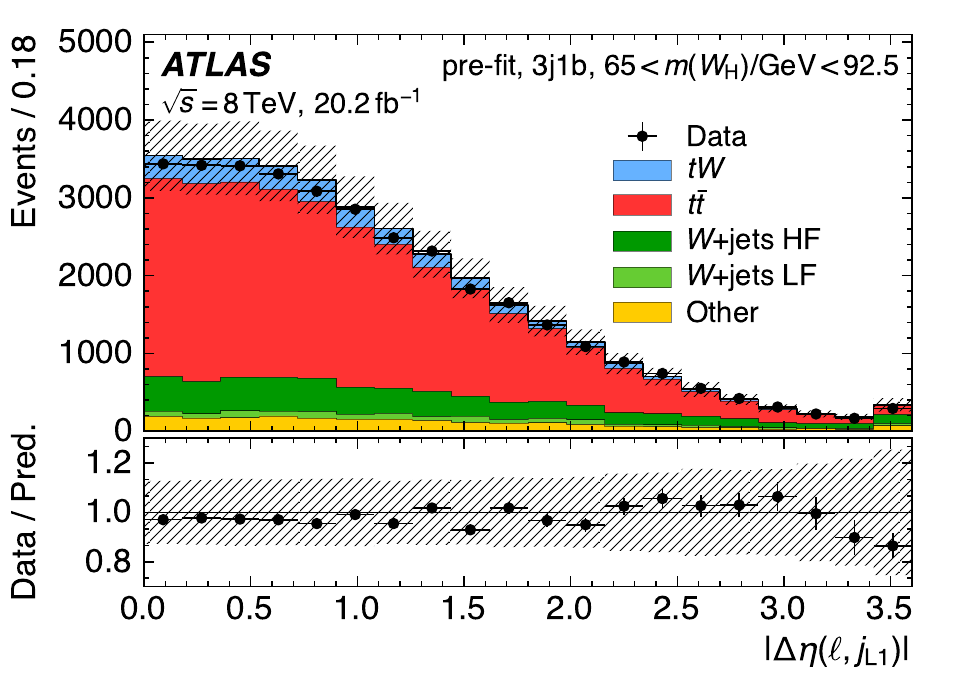}}
\subfloat[]{\includegraphics[scale=0.5]{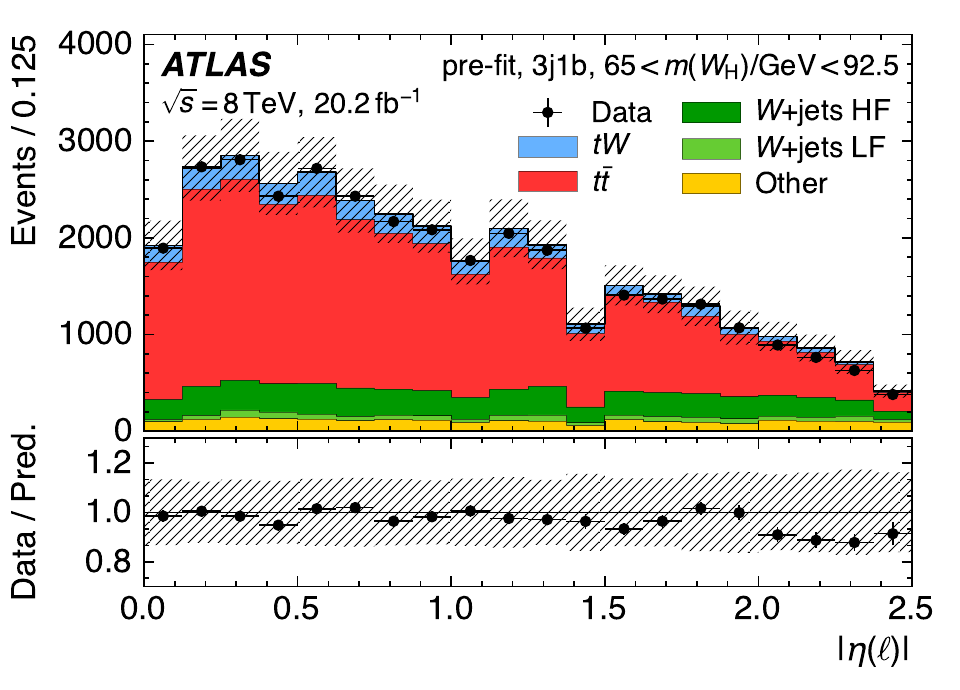}}
\caption{Pre-fit distributions of the NN input variables in the \Wt
signal (3j1b) region with
$\SI{65}{\GeV} \leq \mjj \leq \SI{92.5}{\GeV}$.
\CaptionStack
\CaptionStackOverflow
\CaptionStackRatioS
}
\label{fig:NNVar_Wt}
\end{figure}

\Cref{fig:NNVar_Wt} compares the data with the prediction for the NN input variables.
For all variables, the simulation provides a good description of the data.

\begin{figure}[htbp]
\centering
\subfloat[]{\includegraphics[scale=0.5]{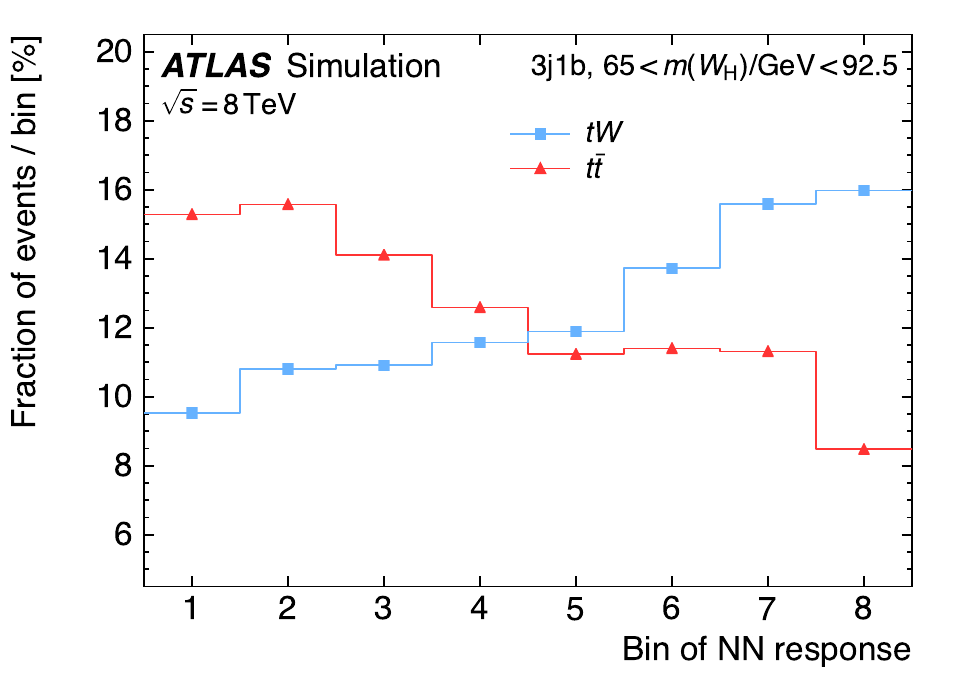}}
\subfloat[]{\includegraphics[scale=0.5]{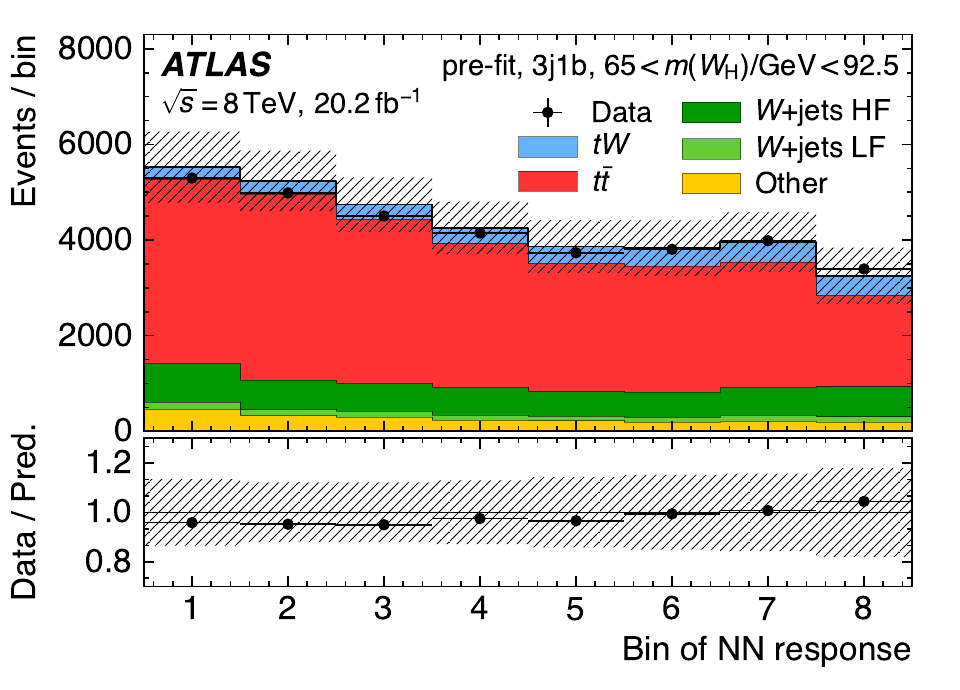}}
\caption{
(a) Shape of the NN response in the signal (3j1b) region.
The distribution contains those events with
$\SI{65}{\GeV}\leq\mjj\leq\SI{92.5}{\GeV}$.
The distributions for the \Wt process and the \ttbar process
normalised to unity are shown.
(b) Pre-fit NN
output distribution in the 3j1b region.
\CaptionStack
\CaptionStackRatio
}
\label{fig:nn-response-binned}
\end{figure}

The distribution of the NN response is subdivided into eight bins, with
the edges placed approximately at the 12.5\% quantiles of a
50:50 mixture of \Wt and \ttbar events.
\Cref{fig:nn-response-binned}(a)
shows the shape of the NN response for the \Wt and \ttbar processes
and \cref{fig:nn-response-binned}(b) presents the
comparison between data and Monte Carlo simulation.

\subsection{Two-dimensional discriminant}
\label{sec:2Ddisc}
 
\begin{figure}[htbp]
\centering
\includegraphics[scale=0.6]{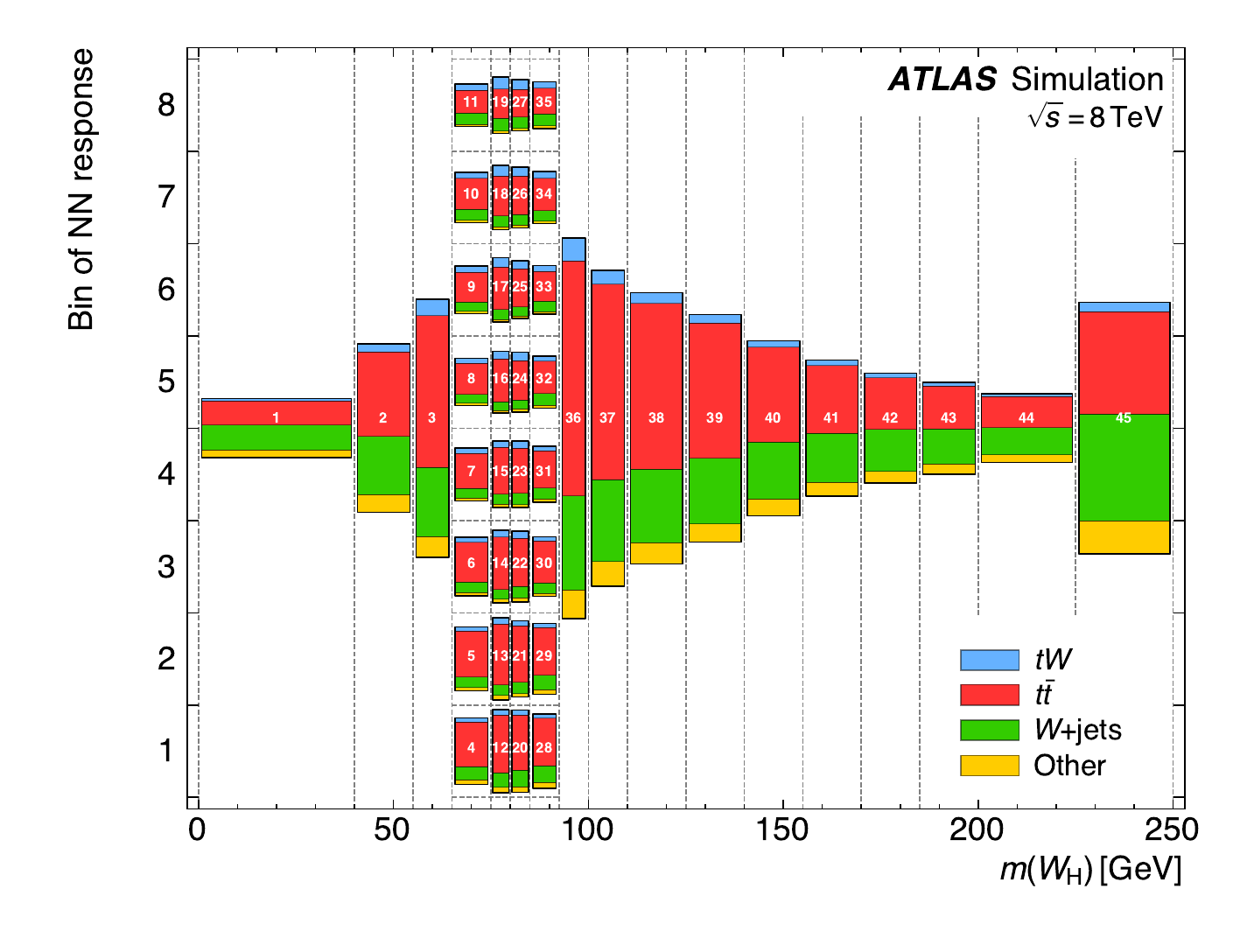}
\caption{
Predicted distribution of the two-dimensional discriminant in the signal (3j1b) region.
The proportions of the coloured areas reflect the expected composition in terms of \Wt, \ttbar, \Wjets and other processes.
The numbers correspond to the bin order when projecting
the discriminant onto one axis as in \Fig{\ref{fig:disco-1d}}.
The last bin on the horizontal axis includes the overflow events.
}
\label{fig:disco-2d}
\end{figure}

For the two-dimensional discriminant, \mjj is used on the abscissa and the NN response on the ordinate of the two-dimensional discriminant.
Outside of the aforementioned \mjj range from \SI{65}{\GeV} to \SI{92.5}{\GeV}, the bins corresponding to different values of the NN response are merged, i.e.\ the NN response is ignored.
The two-dimensional distribution is presented in \Fig{\ref{fig:disco-2d}}.
 
\begin{figure}[htbp]
\centering
\subfloat[]{\includegraphics[scale=0.6]{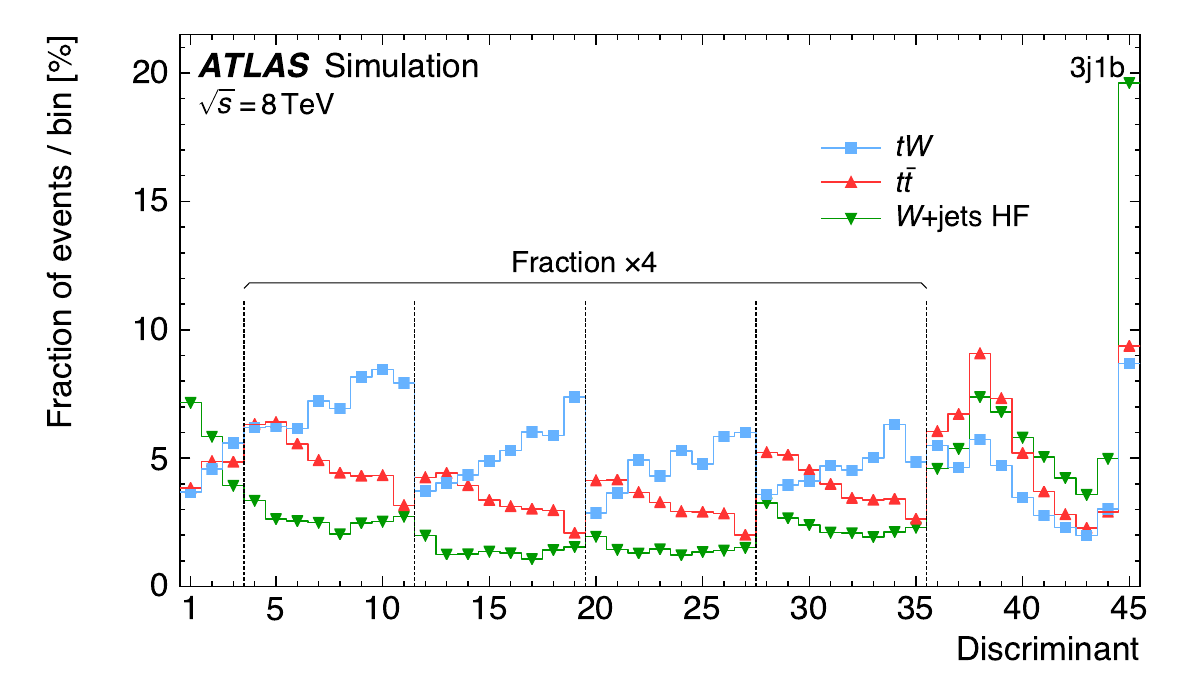}}\\
\subfloat[]{\includegraphics[scale=0.6]{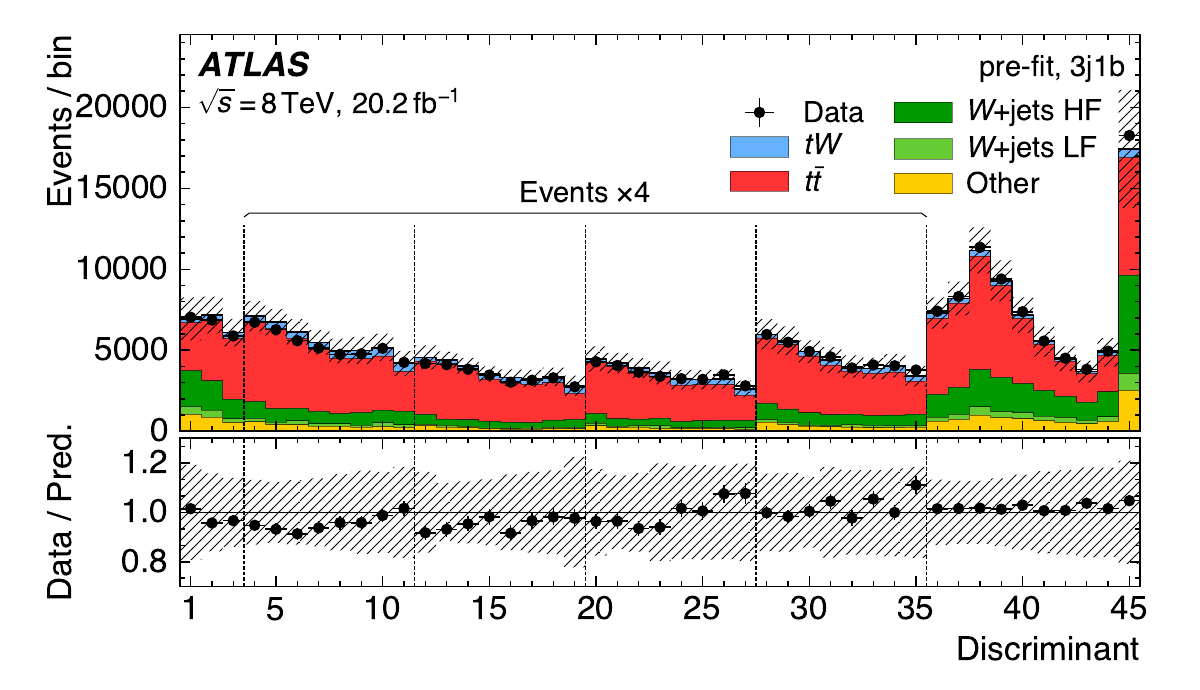}}
\caption{
(a) Shape distribution of
the reconstructed discriminant in the \Wt signal (3j1b) region
rearranged onto a one-dimensional distribution.
\CaptionShape
(b) Pre-fit distributions of
the discriminant in the
\Wt signal (3j1b) region.
\CaptionStack
\CaptionStackRatio{}
The first three bins and the last ten bins correspond directly to (non-uniform) bins of \mjj.
In between are four blocks of eight bins, corresponding to the NN output in slices of \mjj.
Inside each of the blocks, the numbers of events are scaled by a factor of four for better visibility.
}
\label{fig:disco-1d}
\end{figure}

The bins are then rearranged on a one-dimensional axis in column-major order.
The resulting one-dimensional distribution is presented in \Fig{\ref{fig:disco-1d}}, together with a comparison of the shapes.
The first three bins and the last ten bins correspond directly to the
bins of \mjj below \SI{65}{\GeV} and above \SI{92.5}{\GeV} respectively.
In between are four blocks of eight bins, corresponding to the NN output in slices of \mjj.
Inside each of the blocks, the \Wt-to-\ttbar ratio increases significantly from left to right.

\section{Systematic uncertainties}
\label{sec:systematics}
 
Uncertainties in the jet reconstruction arise from the jet energy scale (JES), jet energy resolution (JER), JVF requirement and jet reconstruction efficiency.
The effect of the uncertainty in the JES~\cite{PERF-2014-02} is evaluated by varying the reconstructed energies of the jets in the simulated samples.
It is split into multiple components, taking into account the uncertainty in
the calorimeter response, the detector simulation, the choice of MC event generator, the subtraction of pile-up,
and differences in the detector response for jets initiated by a gluon, a light-flavour quark, or a \Pqb-quark.
In a similar way, the JER uncertainty is represented using several components,
which account for the uncertainty in different \pt and $\eta$ regions of the detector,
the difference between data and MC simulation, as well as the noise contribution in the forward detector region~\cite{PERF-2014-02}.
The uncertainty in jet reconstruction efficiency is estimated by randomly removing simulated jets
from the events according to the jet reconstruction inefficiency measured with dijet events~\cite{PERF-2012-01}.
The JVF uncertainty is evaluated by varying the JVF criterion~\cite{PERF-2014-03}.
 
The scale factors used to correct the \btag efficiency in simulation compared to the efficiency in data
are varied separately for \Pqb-jet, \Pqc-jet and light-flavour jets.
Independent sources of uncertainty affecting the \Pqb-jet tagging efficiency and \Pqc-jet mis-tagging efficiency are considered depending on the jet kinematics,
e.g.\ the variation of the \Pqb-quark jets is subdivided into 6 components.
Uncertainties associated with the lepton selection arise from
the trigger, reconstruction, identification, isolation and lepton momentum scale and resolution~\cite{PERF-2013-03, PERF-2010-04, PERF-2014-05}.
 
All systematic uncertainties in the reconstruction of jets and leptons are propagated to the uncertainty in \met.
In addition, dedicated uncertainties are assigned to the soft term of the \met{},
which accounts for energy deposits in the calorimeter which are not matched to high-\pt physics objects~\cite{PERF-2014-04}.
 
The  uncertainty  in  the  integrated  luminosity  for  the data set used in this analysis is 1.9\%.
It is derived following the methodology detailed in Ref.~\cite{DAPR-2013-01}.
This systematic uncertainty is applied to all contributions determined from the MC simulation.

Uncertainties stemming from theoretical models are evaluated using alternative MC samples for \Wt and \ttbar processes.
The renormalisation and factorisation scales are varied in the matrix element and in the parton shower together with the amount of QCD radiation.
Both scales are varied simultaneously in the matrix element and in the parton shower.
The variation of both \muR and \muF by a factor of \num{0.5} is combined with the \Perugia2012radHi tune,
while the variation of the scale parameters by a factor of \num{2.0} is combined with the \Perugia2012radLo tune~\cite{Skands:2010ak}.
This (radiation) uncertainty is considered uncorrelated between the \Wt and \ttbar processes.
The NLO matrix element generator uncertainty is estimated by comparing two NLO matching methods: \POWHEGBOX and \MCatNLO, both interfaced with \HERWIG{}.
The parton shower, hadronisation and underlying-event systematic uncertainties are computed by comparing \POWHEGBOX with either \PYTHIA or \HERWIG{}.
These are treated as fully correlated between the \Wt and \ttbar processes.
The uncertainty due to the treatment of the interference effects of the \Wt and \ttbar processes is evaluated by using the \Wt DS scheme instead of the DR scheme,
both generated using \POWHEGBOX with \PYTHIA{}.
The effect of the PDF uncertainties on the acceptance is taken into account
for both
the \Wt signal and the \ttbar background
and treated as uncorrelated between the processes,
following the studies in Ref.~\cite{ATL-PHYS-PUB-2015-010}.
 
The uncertainties in the theoretical cross-section calculations are
process dependent and vary from 4\% for the $t$-channel to 6\% for \ttbar (see \Sect{\ref{sec:dataMC}}).
In addition, there are large uncertainties in the $\PZ/\PW\text{+\,jets}$ production cross-sections.
For every jet an additional uncertainty of 24\% is assumed~\cite{Alwall:2007fs}. 
The uncertainty in the normalisation of \PW/\Zboson-boson production in association with three jets is 42\%,
and the rate of \PW-boson events with heavy-flavour jets is allowed to vary by an additional 20\%.
 
The modelling of the \Wjets{} background was cross-checked using \ALPGEN{} with \PYTHIA{}.
The shape of the \Wjets{} background was found to be consistent with the nominal \SHERPA{} prediction.
Hence no dedicated systematic uncertainty is assigned to the choice of generator,
in order to avoid double-counting of the statistical uncertainty of the prediction (\textit{model statistics}).
 
Uncertainties related to the modelling of the fake-lepton background take into account
the choice of control region for the determination of the fake- and real-lepton efficiencies,
the choice of parameterisation, and the normalisation of the prompt-lepton backgrounds in the determination of the efficiencies~\cite{ATLAS-CONF-2014-058}.
 
The uncertainty due to the limited size of the simulated samples
and the fake-lepton background (\textit{model statistics}) is estimated
through the procedure detailed in Refs.~\cite{Barlow:1993dm,Conway:2011in}:
for every bin of the discriminant,
an independent parameter is assigned which describes the variation of the predicted event rate
constrained by its statistical uncertainty.
 
 
 
\section{Statistical analysis}
\label{sec:statanaly}
 
A binned profile maximum-likelihood fit to the discriminant in the signal region is
used to determine the \Wt cross-section.
The likelihood function is defined as a product of Poisson probability terms over all the bins of the discriminant in the signal region and Gaussian penalty terms,
\begin{linenomath}
\begin{equation*}
L(\mu,\vec{\theta};\vec{n}) = \prod_{i}^{\up{bins}} \up{Pois}(n_i;\nu_i(\mu,\vec{\theta})) \, \prod_{k}^{\up{NPs}} G(\theta_k;0,1) \,\text{,}
\end{equation*}
\end{linenomath}
where the $n_i$ ($\nu_i$) is the observed (expected) number of events in each bin $i$ of the discriminant.
The expected number of events depends on the signal-strength
parameter, $\mu$, which is a multiplicative factor on the predicted
signal cross-section.
Nuisance parameters (NPs), $\theta_k$, are used to encode the effects of the systematic uncertainties in the expected number of events.
The Gaussian penalty terms model the external constraints on these parameters.
The estimated parameters, denoted by $\hat{\mu}$ and
$\hat{\theta}$, are obtained by maximising $L(\mu,\vec{\theta};\vec{n})$.
 
The likelihood function is composed and evaluated with the \HistFactory{} program~\cite{CERN-OPEN-2012-016}, part of the \RooStats{} framework~\cite{Moneta:2010pm}.
The minimisation is performed with the \textsc{Minuit} package~\cite{James:1975dr}, using \textsc{Minos} to compute the error estimates.
 
The statistical significance, $Z$, of the result is estimated by comparing the likelihood values of two hypotheses.
The background-only hypothesis is that there is no signal in the data (or equivalently, $\mu=0$).
The signal-plus-background hypothesis is that the signal exists with the signal strength obtained from the fit to data.
With the asymptotic approximation~\cite{Cowan:2010js}, the significance is calculated using a test statistic based on the profile likelihood ratio,
\begin{linenomath}
\begin{equation*}
Z^2 = - 2 \log \frac{
L(\mu=0, \vec\theta=\hat{\vec{\theta}}_{\mu=0}) }{
L(\mu=\hat{\mu}, \vec\theta=\hat{\vec{\theta}}) }\,,
\end{equation*}
\end{linenomath}
where $\hat{\vec{\theta}}_{\mu=0}$ denotes the estimates of the nuisance parameters that maximise the likelihood function under the background-only hypothesis.
The expected significance is calculated by replacing $\vec{n}$ in the likelihood function with the Asimov dataset
for the nominal signal-plus-background hypothesis ($\mu=1$, $\vec{\theta}= \hat{\vec{\theta}}$).
 
 
\section{Cross-section measurement}
\label{sec:fit-result}
 
The \Wt cross-section is extracted from the fit to data in the signal region.
Given the Standard Model prediction, the extracted signal strength is expected to be $\hat{\mu} = \num{1.00(35)}$.
The measured value is $\hat{\mu} = \num{1.16(31)}$,
corresponding to an observed cross-section of $\sigma_{\Wt}^{\up{obs}} =\SI[multi-part-units = single]{26(7)}{\pico\barn}$,
which is consistent with the Standard Model prediction.
The observed (expected) significance is $4.5\sigma$ ($3.9\sigma$).
 
The (post-fit) impact of each systematic uncertainty on the measured signal strength is estimated by means of \textit{conditional} fits, {i.e.} the fit is repeated while keeping the corresponding nuisance parameter fixed at the $\pm1$ standard deviation (sigma) value of the post-fit error interval.
The resulting change in the estimate of the signal strength quantifies the impact of the uncertainty.
For each nuisance parameter, the $+1$ and $-1$ sigma variations are found to be symmetric about the best-fit value to a very good approximation.
Table~\ref{tab:uncertainties} shows the impacts of the systematic uncertainties on the observed fit result,
where the impacts of uncertainties with similar sources have been added in quadrature.
The dominant uncertainties are due to the amount of QCD radiation in signal events and \ttbar{}
background, the JES and \btag, and the model statistics, including the limited size of the MC samples.
 
\begin{table}[htbp]
\centering
\caption{
List of systematic uncertainties considered in the analysis and
their relative impact on the observed signal strength,
evaluated as described in the text.
The \enquote{model statistics} uncertainty is dominated by the \Wjets background.
}
\begin{tabular}{
l
S[table-format=2.0,group-minimum-digits=0]
}
\toprule
\multicolumn{1}{l}{Source} & \multicolumn{1}{c}{Uncertainty [\si{\percent}]} \\
\midrule
Jet energy scale & 10 \\
$b$-tagging & 8 \\
Jet energy resolution & 7 \\
$\met$ reconstruction & 7 \\
Lepton reconstruction & 4 \\
Luminosity & 3 \\
Jet vertex fraction & 3 \\
\midrule
$\ttbar$ radiation & 10 \\
$\Wt$ radiation & 9 \\
\Wt--\ttbar interference & 7 \\
\ttbar cross-section normalisation & 6 \\
Other background cross-section normalisations & 5 \\
$\Wt$ and $\ttbar$ parton shower & 4 \\
$\Wt$ and $\ttbar$ NLO matching & 3 \\
PDF & 1 \\
\midrule
Model statistics & 11 \\
Data statistics & 4 \\
\midrule
Total & 27 \\
\bottomrule
\end{tabular}
\label{tab:uncertainties}
\end{table}
 
Some nuisance parameters are constrained by the
data. For example, the normalisation uncertainty for \Wjets events is
reduced from 45\% to 8\%, because the assigned initial uncertainty is large
and this background can be separated well from \Wt and \ttbar events.
By design of the discriminant, combinations of nuisance parameters
that shift the peak in the \mjj distribution are constrained,
primarily the JES and choice of renormalisation scale together with the amount of
QCD radiation in signal and \ttbar{} background.
Also, the nuisance parameter for the NLO matching for \Wt and \ttbar is constrained:
the choice of \MCatNLO{} is not supported by the data,
reducing the impact of the choice from 9\% pre-fit to 3\% post-fit.
 
\begin{table}[hbp]
\centering
\caption{
Post-fit signal and background and observed number of events in
the signal region and the \ttbar validation region.
The uncertainties include statistical plus all systematic uncertainties  (cf.\ \Sect{\ref{sec:systematics}}).
}
\setlength{\tabcolsep}{3pt}
\begin{tabular}{
l
S[table-format=6.0,group-minimum-digits=0]@{$\,\pm\,$}
S[table-format=4.0,group-minimum-digits=0]
S[table-format=5.0,group-minimum-digits=0]@{$\,\pm\,$}
S[table-format=4.0,group-minimum-digits=0]
}
\toprule
\multicolumn{1}{c}{Process} & \multicolumn{2}{c}{Signal region} & \multicolumn{2}{c}{\ttbar{} region} \\
\multicolumn{1}{c}{} & \multicolumn{2}{c}{(3j1b)} & \multicolumn{2}{c}{(4j2b)}\\
\midrule
\Wt{} & 7800 & 1800 & 1300 & 400 \\
\midrule
\ttbar{} & 74500 & 2100 & 36700 & 2300 \\
$t$, $t$-channel & 4250 & 200 & 590 & 40 \\
$t$, $s$-channel & 315 & 15 & 63 & 4 \\
\Wjets{}, HF & 34700 & 3300 & 1400 & 500\\
\Wjets{}, other & 5700 & 1800 & 27 & 19 \\
\Zjets{} & 3800 & 1500 & 180 & 90\\
\VVjets{} & 640 & 270 & 23 & 13 \\
Fake leptons & 3000 & 1600 & 5 & 22 \\
\midrule
Total background & 126900 & 1900 & 38900 & 2400 \\
\midrule
Total signal + background & 134700 & 500 & 40200 & 2300 \\
Observed & \multicolumn{1}{S[table-format=6.0,group-minimum-digits=0]@{\phantom{$\,\pm\,$}}}{134633} &  & \multicolumn{1}{S[table-format=5.0,group-minimum-digits=0]@{\phantom{$\,\pm\,$}}}{41738} &   \\
\bottomrule
\end{tabular}
\label{tab:postfit-yields_no4j0b}
\end{table}
 
A few nuisance parameters are pulled away from the pre-fit expectation.
For the parameter associated with the choice of parton-shower generator, a blend of \PYTHIA{} and \HERWIG{} gives the best description of the data, while the nominal \PYTHIA{} prediction is disfavoured at the two-sigma level.
The $b$-tagging parameter with the largest effect on the overall $b$-tagging efficiency
is pulled by about one sigma, corresponding to a
decrease of about 1\% to 2\% in the $b$-tagging efficiency compared to the pre-fit expectation.
Given that the $b$-tagging calibration partially relies on dijet events~\cite{PERF-2012-04},
which correspond to a different environment regarding the production mechanism of the $b$-jets, the pull is reasonable.
 
\Cref{tab:postfit-yields_no4j0b} shows the post-fit event yields of each
process.
The uncertainties in the yields are computed taking the correlations between nuisance parameters and processes into account.
The post-fit estimates are well within the uncertainties of the pre-fit expectation (\cref{tab:prefit-yields_onlytW}),
while most of their uncertainties are reduced.
The normalisation uncertainty for \PW\text{+\,HF jets} changes from almost
50\% to about 10\%.
 
\begin{figure}[htbp]
\centering
\subfloat[]{\includegraphics[scale=0.5]{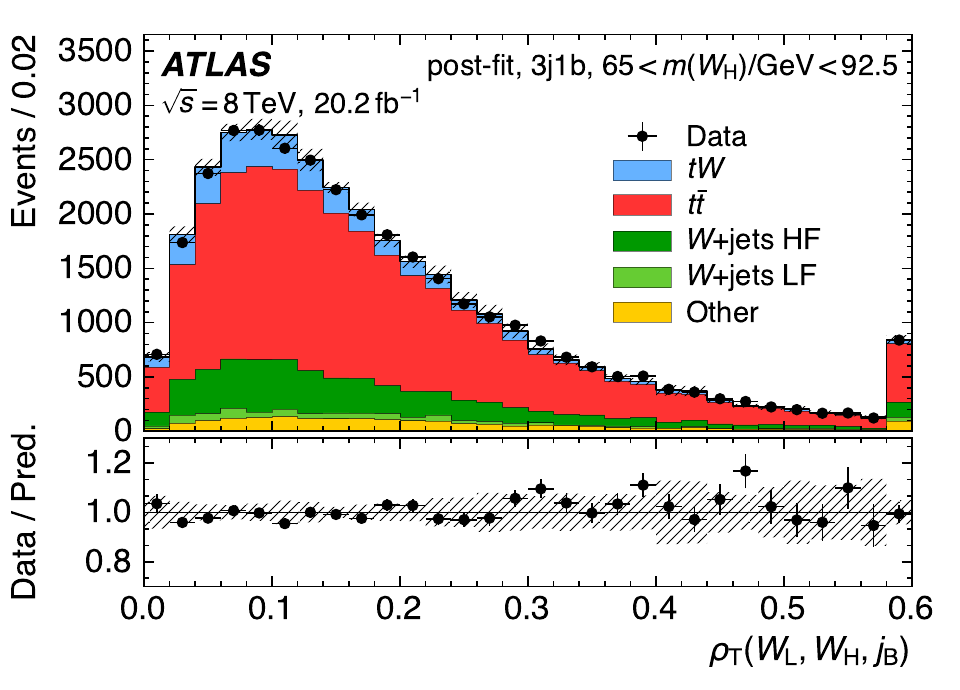}}  
\subfloat[]{\includegraphics[scale=0.5]{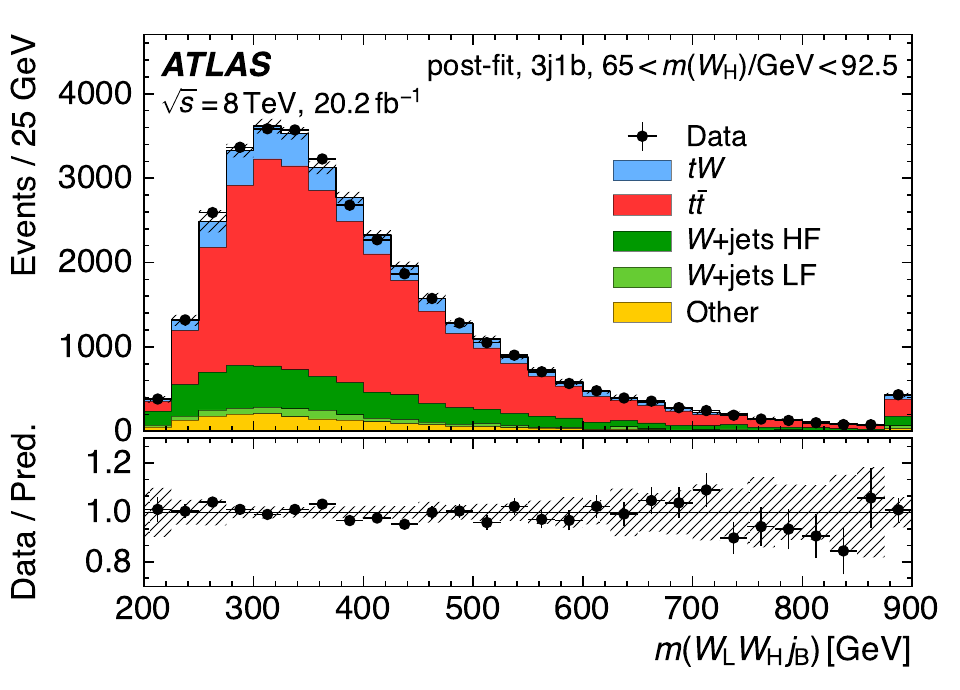}}\\ 
\subfloat[]{\includegraphics[scale=0.5]{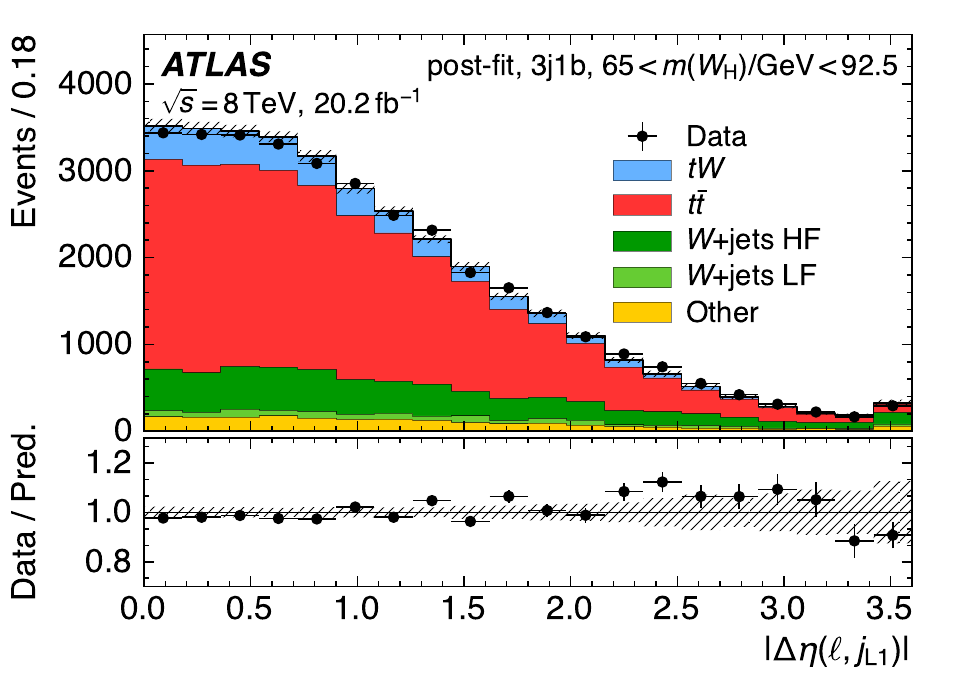}}
\subfloat[]{\includegraphics[scale=0.5]{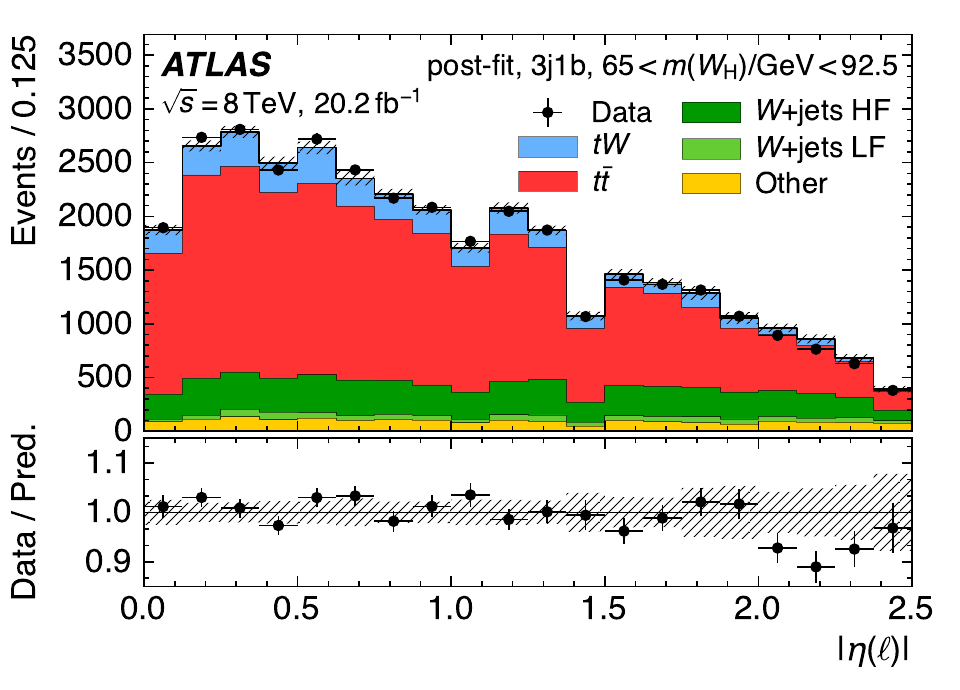}}\\ 
\subfloat[]{\includegraphics[scale=0.5]{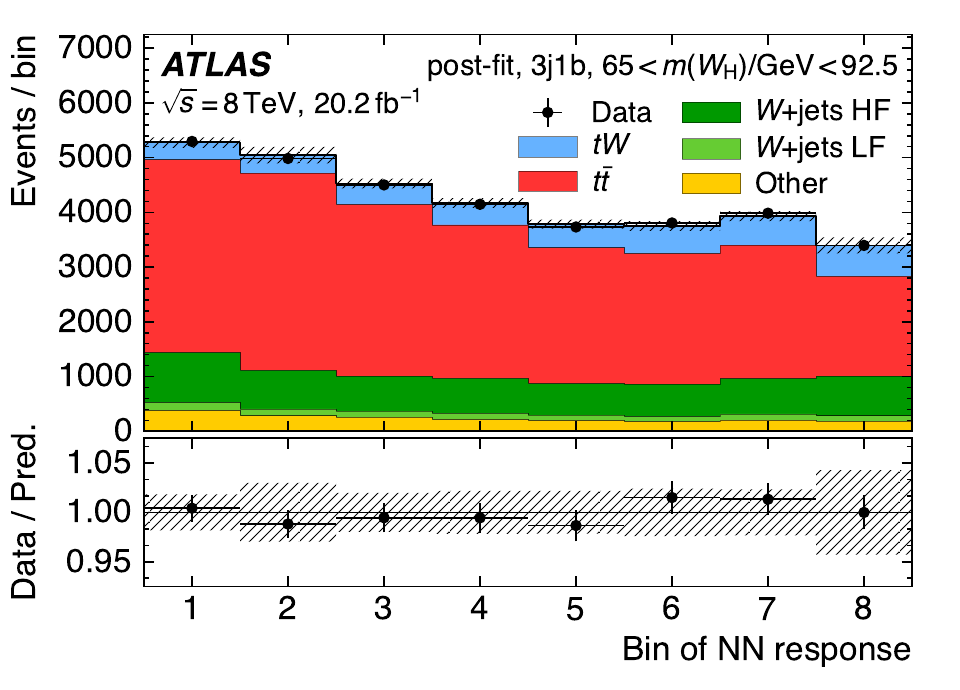}} 
\subfloat[]{\includegraphics[scale=0.5]{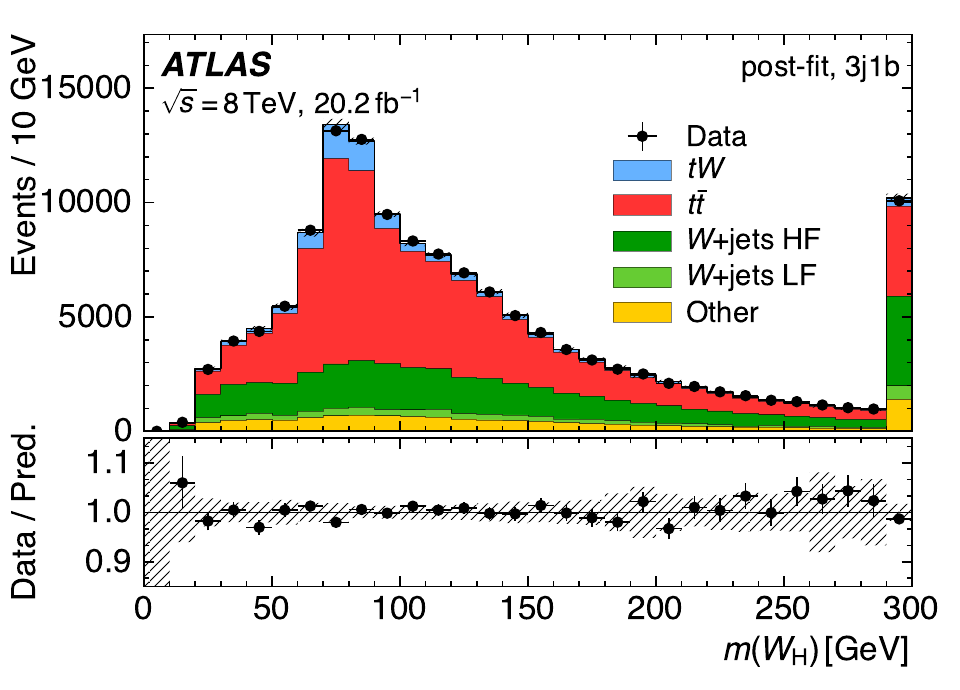}} 
\caption{
(a), (b), (c), (d) Post-fit distributions of the NN input variables,
(e) NN discriminant
and (f) \mjj in the signal region.
\CaptionStackPostFit{}
The last bin includes the overflow events, except for (e).
\CaptionStackRatioS
}
\label{fig:postfit-NN-MW}
\end{figure}

\Cref{fig:postfit-NN-MW} shows the  post-fit distributions for the NN
input variables, the NN output response and the \mjj in the signal
region.
The post-fit plots use the parameter estimates obtained in the fit of
the discriminant, including their uncertainties,
and demonstrate a good description of the data.
 
\begin{figure}[htbp]
\centering
\subfloat[]{\includegraphics[scale=0.5]{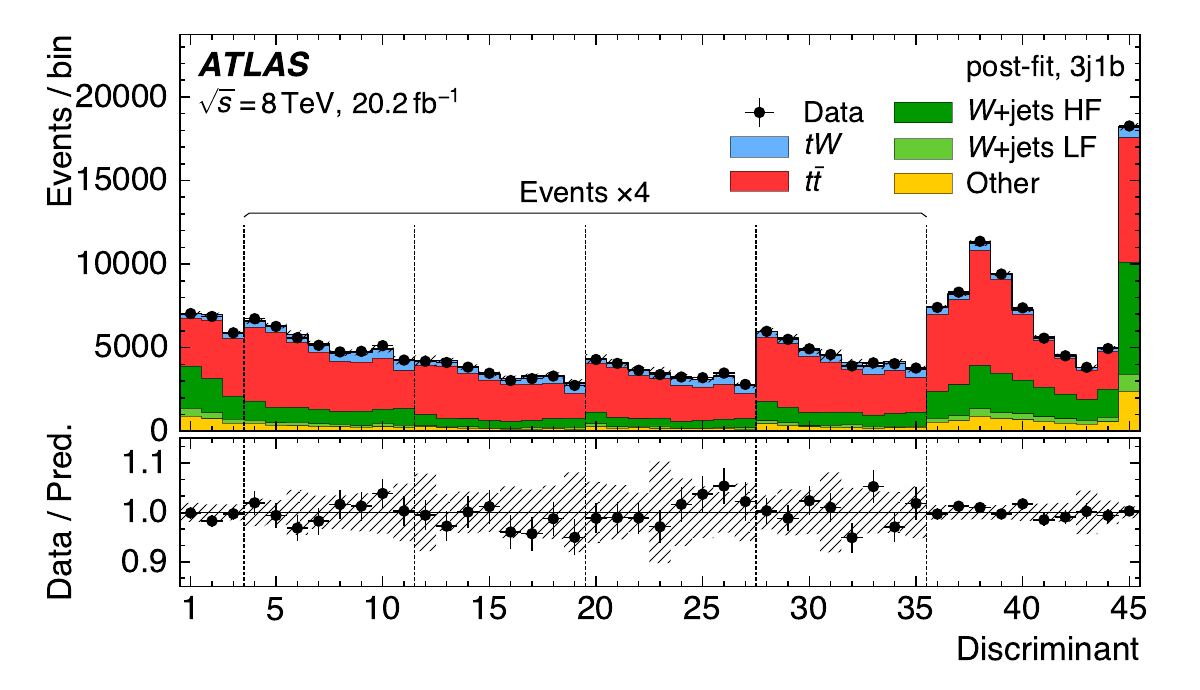}}\\
\subfloat[]{\includegraphics[scale=0.5]{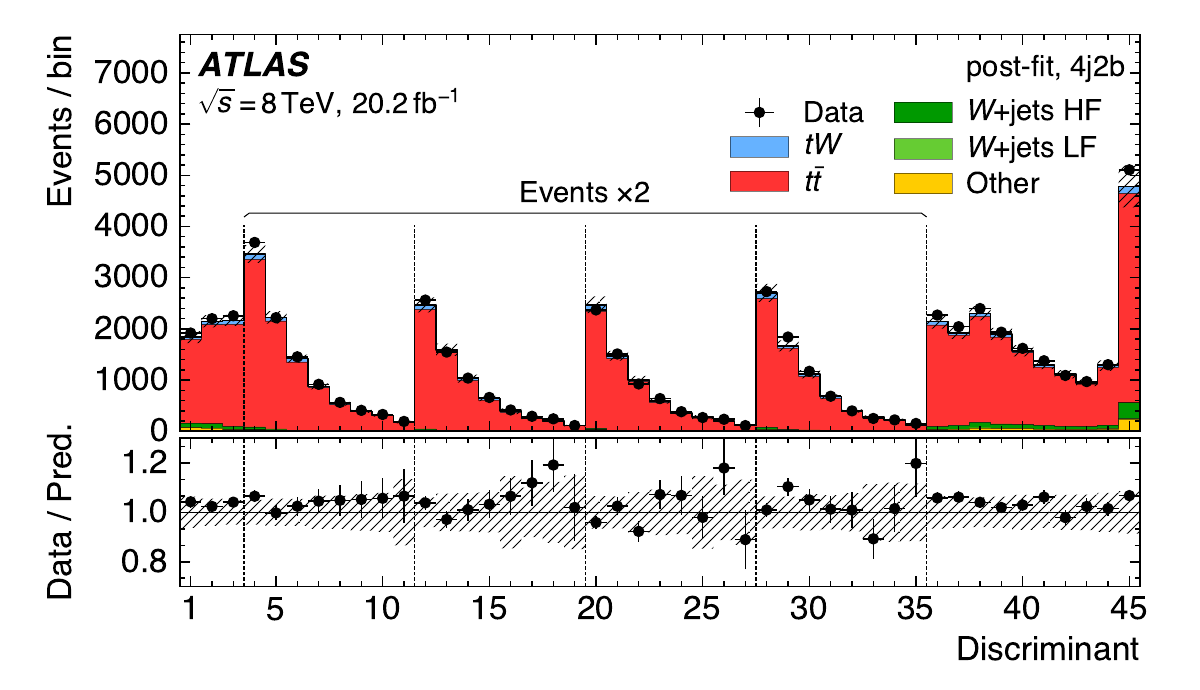}}
\caption{
Post-fit distributions of the discriminant in the
(a) signal region and
(b) validation region.
\CaptionStackPostFit
\CaptionStackRatioS{}
The first three bins and the last ten bins correspond directly to (non-uniform) bins of \mjj.
In between are four blocks of eight bins, corresponding to the NN output in slices of \mjj.
Inside each of the blocks, the numbers of events are scaled by a factor of four (factor of two in 4j2b) for better visibility.
}
\label{fig:postfit-disco}
\end{figure}

\Cref{fig:postfit-disco}(a) shows that the data are well described by the model
in the signal region.
\Cref{fig:postfit-disco}(b) shows
the strongest support for the validity of the fit result by comparing
the expected distributions and observed distributions in the
\ttbar validation region.
It shows that
the uncertainty due to the extrapolation from the signal region is small,
and therefore provides a stringent test that the main background is well modelled.

\FloatBarrier
 
\section{Conclusion}
\label{sec:conclusion}
The inclusive cross-section for the production of a single top quark
in association with a \PW boson in the
single-lepton channel  is measured using an integrated
luminosity of \SI{20.2}{\per\femto\barn} of $\sqrt{s}=\SI{8}{\TeV}$
proton--proton collision data collected by
the ATLAS detector at the LHC in 2012.
A neural network is used to separate the signal from the \ttbar background.
A two-dimensional discriminant, built from the
neural-network response and the mass of the hadronically decaying \PW boson, is used to extract
the cross-section.
Evidence for \Wt production in the single-lepton channel is obtained with
an observed (expected) significance of $4.5$  ($3.9$) standard
deviations.
The measured cross-section is:
\begin{linenomath}
\begin{equation*}
\sigma_{\Wt }^{\up{meas.}} = \SI[multi-part-units = single]{26(7)}{\pico\barn}\,,
\end{equation*}
\end{linenomath}
which is consistent with the SM expectation of
$\sigma_{\Wt}^{\text{th.}}
= \SI[multi-part-units = single]{22.4(15)}{\pico\barn}$.


\section*{Acknowledgements}

We thank CERN for the very successful operation of the LHC, as well as the
support staff from our institutions without whom ATLAS could not be
operated efficiently.
 
We acknowledge the support of ANPCyT, Argentina; YerPhI, Armenia; ARC, Australia; BMWFW and FWF, Austria; ANAS, Azerbaijan; SSTC, Belarus; CNPq and FAPESP, Brazil; NSERC, NRC and CFI, Canada; CERN; ANID, Chile; CAS, MOST and NSFC, China; COLCIENCIAS, Colombia; MSMT CR, MPO CR and VSC CR, Czech Republic; DNRF and DNSRC, Denmark; IN2P3-CNRS and CEA-DRF/IRFU, France; SRNSFG, Georgia; BMBF, HGF and MPG, Germany; GSRT, Greece; RGC and Hong Kong SAR, China; ISF and Benoziyo Center, Israel; INFN, Italy; MEXT and JSPS, Japan; CNRST, Morocco; NWO, Netherlands; RCN, Norway; MNiSW and NCN, Poland; FCT, Portugal; MNE/IFA, Romania; JINR; MES of Russia and NRC KI, Russian Federation; MESTD, Serbia; MSSR, Slovakia; ARRS and MIZ\v{S}, Slovenia; DST/NRF, South Africa; MICINN, Spain; SRC and Wallenberg Foundation, Sweden; SERI, SNSF and Cantons of Bern and Geneva, Switzerland; MOST, Taiwan; TAEK, Turkey; STFC, United Kingdom; DOE and NSF, United States of America. In addition, individual groups and members have received support from BCKDF, CANARIE, Compute Canada, CRC and IVADO, Canada; Beijing Municipal Science \& Technology Commission, China; COST, ERC, ERDF, Horizon 2020 and Marie Sk{\l}odowska-Curie Actions, European Union; Investissements d'Avenir Labex, Investissements d'Avenir Idex and ANR, France; DFG and AvH Foundation, Germany; Herakleitos, Thales and Aristeia programmes co-financed by EU-ESF and the Greek NSRF, Greece; BSF-NSF and GIF, Israel; La Caixa Banking Foundation, CERCA Programme Generalitat de Catalunya and PROMETEO and GenT Programmes Generalitat Valenciana, Spain; G\"{o}ran Gustafssons Stiftelse, Sweden; The Royal Society and Leverhulme Trust, United Kingdom.
 
The crucial computing support from all WLCG partners is acknowledged gratefully, in particular from CERN, the ATLAS Tier-1 facilities at TRIUMF (Canada), NDGF (Denmark, Norway, Sweden), CC-IN2P3 (France), KIT/GridKA (Germany), INFN-CNAF (Italy), NL-T1 (Netherlands), PIC (Spain), ASGC (Taiwan), RAL (UK) and BNL (USA), the Tier-2 facilities worldwide and large non-WLCG resource providers. Major contributors of computing resources are listed in Ref.~\cite{ATL-SOFT-PUB-2020-001}.
 

\printbibliography

@article{Tait:2000sh,
      author         = "Tait, Tim M. P. and Yuan, C.-P.",
      title          = "{Single top quark production as a window to physics
                        beyond the standard model}",
      journal        = "Phys. Rev. D",
      volume         = "63",
      year           = "2000",
      pages          = "014018",
      doi            = "10.1103/PhysRevD.63.014018",
      eprint         = "hep-ph/0007298",
      archivePrefix  = "arXiv",
      reportNumber   = "ANL-HEP-PR-98-124, CERN-TH-2000-224",
}

@Article{Martin2009,
     author="Martin, A. D. and Stirling, W. J. and Thorne, R. S. and Watt, G.",
     title="Parton distributions for the LHC ",
     journal="Eur. Phys. J. C",
     year="2009",
     volume="63",
     number="2",
     pages="189--285",
     issn="1434-6052",
     doi="10.1140/epjc/s10052-009-1072-5",
     eprint         = "0901.0002",
     archivePrefix  = "arXiv",
     primaryClass   = "hep-ph",     
}

@article{Sjostrand:1064095,
      author        = "Sjöstrand, Torbjörn and Mrenna, Stephen and Skands,
                       Peter",
      title         = "{A brief introduction to PYTHIA 8.1}",
      journal       = "Comput. Phys. Commun.",
      doi           = "10.1016/j.cpc.2008.01.036",
      volume        = "178",
      pages         = "852-867",
      month         = 10,
      year          = "2007",
      eprint         = "0710.3820",
      archivePrefix  = "arXiv",
      primaryClass   = "hep-ph",
}

@article{Nadolsky:2008zw,
      author         = "Nadolsky, Pavel M. and Lai, Hung-Liang and Cao, Qing-Hong
                        and Huston, Joey and Pumplin, Jon and Stump, Daniel and
                        Tung, Wu-Ki and Yuan, C.-P.",
      title          = "{Implications of CTEQ global analysis for collider
                        observables}",
      journal        = "Phys. Rev. D",
      volume         = "78",
      year           = "2008",
      pages          = "013004",
      doi            = "10.1103/PhysRevD.78.013004",
      eprint         = "0802.0007",
      archivePrefix  = "arXiv",
      primaryClass   = "hep-ph",
      reportNumber   = "MSUHEP-080125, UCRHEP-T447",
      SLACcitation   = "%%CITATION = ARXIV:0802.0007;%%"
}

@booklet{Botje:1318947,
      author        = "Botje, Michiel and Butterworth, Jon and Cooper-Sarkar,
                       Amanda and de Roeck, Albert and Feltesse, Joel and Forte,
                       Stefano and Glazov, Alexander and Huston, Joey and McNulty,
                       Ronan and Sjostrand, Torbjorn and Thorne, Robert",
      title         = "{The PDF4LHC Working Group Interim Recommendations}",
      number        = "arXiv:1101.0538",
      month         = "1",
      year          = "2011",
      reportNumber  = "arXiv:1101.0538",
      url           = "https://cds.cern.ch/record/1318947",
}

@Article{Alwall:2007fs,
      author         = "Alwall, J. and others",
      title          = "{Comparative study of various algorithms for the merging of parton showers and matrix elements in hadronic collisions}",
      journal        = "Eur. Phys. J. C",
      volume         = "53",
      year           = "2008",
      pages          = "473",
      doi            = "10.1140/epjc/s10052-007-0490-5",
      eprint         = "0706.2569",
      archivePrefix  = "arXiv",
      primaryClass   = "hep-ph",
}

@article{PhysRevD.69.094008,
  author = "Anastasiou, Charalampos and Dixon, Lance and Melnikov, Kirill and Petriello, Frank",
  title = "{High-precision QCD at hadron colliders: Electroweak gauge boson rapidity distributions at next-to-next-to leading order}",
  journal = {Phys. Rev. D},
  volume = {69},
  pages = {094008},
  year = {2004},
  doi = {10.1103/PhysRevD.69.094008},
  archivePrefix = "arXiv",
  eprint = "hep-ph/0312266",
}

@article{Campbell:1999ah,
    author = "Campbell, John M. and Ellis, R.Keith",
    title = "{An Update on vector boson pair production at hadron colliders}",
    reportNumber = "FERMILAB-PUB-99-146-T",
    journal = "Phys. Rev. D",
    volume = "60",
    year = "1999",
    pages = "113006",
    doi = "10.1103/PhysRevD.60.113006",
    archivePrefix = "arXiv",
    eprint = "hep-ph/9905386",
}

@article{Cao:2007ea,
      author         = "Cao, Qing-Hong and Wudka, Jose and Yuan, C.-P.",
      title          = "{Search for new physics via single-top production at the
                        LHC}",
      journal        = "Phys. Lett. B",
      volume         = "658",
      year           = "2007",
      pages          = "50-56",
      doi            = "10.1016/j.physletb.2007.10.057",
      eprint         = "0704.2809",
      archivePrefix  = "arXiv",
      primaryClass   = "hep-ph",
      reportNumber   = "UCRHEP-T427",
}

@Article{Agostinelli:2002hh,
     author          = "Agostinelli, S. and others",
     collaboration   = "GEANT4",
     title           = "{GEANT4 -- a simulation toolkit}",
     journal         = "Nucl. Instrum. Meth. A",
     volume          = "506",
     year            = "2003",
     pages           = "250-303",
     doi             = "10.1016/S0168-9002(03)01368-8",
}

@article{1126-6708-2006-05-026,
  author={Sjöstrand, Torbjörn  and Mrenna, Stephen and Skands, Peter},
  title={PYTHIA 6.4 physics and manual},
  journal={JHEP},
  volume={05},
  pages={026},
  doi = "10.1088/1126-6708/2006/05/026",
  year={2006},
  eprint         = "hep-ph/0603175",
  archivePrefix  = "arXiv",
  reportNumber   = "FERMILAB-PUB-06-052-CD-T, LU-TP-06-13",
}

@article{1126-6708-2007-11-070,
  author={Stefano Frixione and Paolo Nason and Carlo Oleari},
  title={Matching NLO QCD computations with parton shower simulations: the POWHEG method},
  journal={JHEP},
  volume={11},
  number={11},
  pages={070},
  doi = "10.1088/1126-6708/2007/11/070",
  year={2007},
  eprint         = "0709.2092",
  archivePrefix  = "arXiv",
  primaryClass   = "hep-ph",
  reportNumber   = "BICOCCA-FT-07-9, GEF-TH-21-2007",
}

@article{1126-6708-2004-11-040,
  author={Paolo Nason},
  title={A new method for combining NLO QCD with shower Monte Carlo algorithms},
  journal={JHEP},
  volume={11},
  number={11},
  pages={040},
  doi = "10.1088/1126-6708/2004/11/040",
  year={2004},
  eprint         = "hep-ph/0409146",
  archivePrefix  = "arXiv",
  reportNumber   = "BICOCCA-FT-04-11",
}

@Article{Re:2010bp,
      author         = "Re, Emanuele",
      title          = "{Single-top \(Wt\)-channel production matched with parton showers using the POWHEG method}",
      journal        = "Eur. Phys. J. C",
      volume         = "71",
      year           = "2011",
      pages          = "1547",
      doi            = "10.1140/epjc/s10052-011-1547-z",
      eprint         = "1009.2450",
      archivePrefix  = "arXiv",
      primaryClass   = "hep-ph",
      reportNumber   = "IPPP-10-74, DCPT-10-148",
      SLACcitation   = "%%CITATION = ARXIV:1009.2450;%%"
}

@Article{Alioli2010,
     author="Alioli, Simone
     and Nason, Paolo
     and Oleari, Carlo
     and Re, Emanuele",
     title="A general framework for implementing NLO calculations in shower Monte Carlo programs: the POWHEG BOX",
     journal="JHEP",
     year="2010",
     volume="06",
     pages="1--58",
     abstract="In this work we illustrate the POWHEG BOX, a general computer code framework for implementing NLO calculations in shower Monte Carlo programs according to the POWHEG method. Aim of this work is to provide an illustration of the needed theoretical ingredients, a view of how the code is organized and a description of what a user should provide in order to use it.",
     %issn="1029-8479",
     doi="10.1007/JHEP06(2010)043",
     eprint         = "1002.2581",
     archivePrefix  = "arXiv",
     primaryClass   = "hep-ph",
}

@article{PhysRevD.82.074024,
  title = {New parton distributions for collider physics},
  author = {Lai, Hung-Liang and Guzzi, Marco and Huston, Joey and Li, Zhao and Nadolsky, Pavel M. and Pumplin, Jon and Yuan, C.-P.},
  journal = {Phys. Rev. D},
  volume = {82},
  %issue = {7},
  pages = {074024},
  numpages = {24},
  year = {2010},
  month = 10,
  publisher = {American Physical Society},
  doi = {10.1103/PhysRevD.82.074024},
  eprint         = "1007.2241",
  archivePrefix  = "arXiv",
  primaryClass   = "hep-ph",
  reportNumber   = "MSUHEP-100707, SMU-HEP-10-10",
}

@article{Skands:2010ak,
      author         = "Skands, Peter Zeiler",
      title          = "{Tuning Monte Carlo generators: The Perugia tunes}",
      journal        = "Phys. Rev. D",
      volume         = "82",
      year           = "2010",
      pages          = "074018",
      doi            = "10.1103/PhysRevD.82.074018",
      eprint         = "1005.3457",
      archivePrefix  = "arXiv",
      primaryClass   = "hep-ph",
      reportNumber   = "MCNET-10-08, CERN-PH-TH-2010-113",
      SLACcitation   = "%%CITATION = ARXIV:1005.3457;%%"
}

@article{Frixione:2008yi,
      author         = "Frixione, Stefano and Laenen, Eric and Motylinski,
                        Patrick and Webber, Bryan R. and White, Chris D.",
      title          = "{Single-top hadroproduction in association with a $W$
                        boson}",
      journal        = "JHEP",
      volume         = "07",
      year           = "2008",
      pages          = "029",
      doi            = "10.1088/1126-6708/2008/07/029",
      eprint         = "0805.3067",
      archivePrefix  = "arXiv",
      primaryClass   = "hep-ph",
      reportNumber   = "CERN-TH-2008-104, ITP-UU-08-27, NIKHEF-2008-005,
                        ITFA-2008-16, CAVENDISH-HEP-08-06",
}

@article{White:2009yt,
      author         = "White, Chris D. and Frixione, Stefano and Laenen, Eric
                        and Maltoni, Fabio",
      title          = "Isolating $Wt$ production at the {LHC}",
      journal        = "JHEP",
      volume         = "11",
      pages          = "074",
      doi            = "10.1088/1126-6708/2009/11/074",
      year           = "2009",
      eprint         = "0908.0631",
      archivePrefix  = "arXiv",
      primaryClass   = "hep-ph",
      reportNumber   = "NIKHEF-2009-013, CERN-TH-2009-136, ITF-UU-09-31, CP3-09-33",
}

@article{1126-6708-2002-06-029,
  author={Frixione, Stefano and Webber, Bryan R.},
  title={Matching NLO QCD computations and parton shower simulations},
  journal={JHEP},
  volume={06},
  number={06},
  pages={029},
  doi = "10.1088/1126-6708/2002/06/029",
  year={2002},
  eprint         = "hep-ph/0204244",
  archivePrefix  = "arXiv",
  reportNumber   = "CAVENDISH-HEP-02-01, LAPTH-905-02, GEF-TH-2-2002",
}

@article{1126-6708-2001-01-010,
  author={Gennaro Corcella and Ian G. Knowles and Giuseppe Marchesini and Stefano Moretti and Kosuke Odagiri and Peter
  Richardson and Michael H. Seymour and Bryan R. Webber},
  title={HERWIG 6: an event generator for hadron emission reactions with interfering gluons (including supersymmetric processes)},
  journal={JHEP},
  volume={01},
  number={01},
  pages={010},
  doi = "10.1088/1126-6708/2001/01/010",
  year={2001},
  eprint         = "hep-ph/0011363",
  archivePrefix  = "arXiv",
  reportNumber   = "CAVENDISH-HEP-99-03, CERN-TH-2000-284, RAL-TR-2000-048",
}

@Article{Butterworth1996,
    author="Butterworth, J. M. and Forshaw, J. R. and Seymour, M. H.",
    title="Multiparton interactions in photoproduction at {HERA}",
    journal="Z. f{\"u}r Physik C",
    year="1996",
    volume="72",
    number="4",
    pages="637--646",
    doi="10.1007/s002880050286",
    eprint         = "hep-ph/9601371",
    archivePrefix  = "arXiv",
    reportNumber   = "CERN-TH-96-82, M-C-TH-96-05, UCL-HEP-96-02"
}

@article{Kidonakis:2010ux,
      author         = "Kidonakis, Nikolaos",
      title          = "{Two-loop soft anomalous dimensions for single top quark
                        associated production with a \(W^{-}\) or \(H^{-}\)}",
      journal        = "Phys. Rev. D",
      volume         = "82",
      year           = "2010",
      pages          = "054018",
      doi            = "10.1103/PhysRevD.82.054018",
      eprint         = "1005.4451",
      archivePrefix  = "arXiv",
      primaryClass   = "hep-ph",
      SLACcitation   = "%%CITATION = ARXIV:1005.4451;%%"
}

@Inproceedings{Kidonakis:2013zqa,
      author         = "Kidonakis, Nikolaos",
      title          = "{Top Quark Production}",
      booktitle      = "{Proceedings, Helmholtz International Summer School on
                        Physics of Heavy Quarks and Hadrons (HQ 2013): JINR,
                        Dubna, Russia, July 15-28, 2013}",
      year           = "2014",
      pages          = "139-168",
      doi            = "10.3204/DESY-PROC-2013-03/Kidonakis",
      eprint         = "1311.0283",
      archivePrefix  = "arXiv",
      primaryClass   = "hep-ph",
      reportNumber   = "KSU-HEP-110113",
      SLACcitation   = "%%CITATION = ARXIV:1311.0283;%%"
}

@Article{Frixione:2007nw,
      author         = "Frixione, Stefano and Nason, Paolo and Ridolfi, Giovanni",
      title          = "{A positive-weight next-to-leading-order Monte Carlo for
                        heavy flavour hadroproduction}",
      journal        = "JHEP",
      volume         = "09",
      year           = "2007",
      pages          = "126",
      doi            = "10.1088/1126-6708/2007/09/126",
      eprint         = "0707.3088",
      archivePrefix  = "arXiv",
      primaryClass   = "hep-ph",
      reportNumber   = "BICOCCA-FT-07-12, GEF-TH-19-2007",
      SLACcitation   = "%%CITATION = ARXIV:0707.3088;%%"
}

@article{Czakon:2011xx,
      author         = "Czakon, Michal and Mitov, Alexander",
      title          = "{Top++: A program for the calculation of the top-pair
                        cross-section at hadron colliders}",
      journal        = "Comput. Phys. Commun.",
      volume         = "185",
      year           = "2014",
      pages          = "2930",
      doi            = "10.1016/j.cpc.2014.06.021",
      eprint         = "1112.5675",
      archivePrefix  = "arXiv",
      primaryClass   = "hep-ph",
      reportNumber   = "CERN-PH-TH-2011-303, TTK-11-58",
      SLACcitation   = "%%CITATION = ARXIV:1112.5675;%%"
}

@article{Demartin:2010er,
      author         = "Demartin, Federico and Forte, Stefano and Mariani, Elisa
                        and Rojo, Juan and Vicini, Alessandro",
      title          = "{Impact of parton distribution function and $\alpha_{s}$ uncertainties on Higgs boson production in gluon fusion at hadron colliders}",
      journal        = "Phys. Rev. D",
      volume         = "82",
      year           = "2010",
      pages          = "014002",
      doi            = "10.1103/PhysRevD.82.014002",
      eprint         = "1004.0962",
      archivePrefix  = "arXiv",
      primaryClass   = "hep-ph",
      reportNumber   = "IFUM-956-FT",
      SLACcitation   = "%%CITATION = ARXIV:1004.0962;%%"
}

@article{PhysRevD.89.033009,
  title = {CT10 next-to-next-to-leading order global analysis of QCD},
  author = {Gao, Jun and Guzzi, Marco and Huston, Joey and Lai, Hung-Liang and Li, Zhao and Nadolsky, Pavel and Pumplin, Jon and Stump, Daniel and Yuan, C.-P.},
  journal = {Phys. Rev. D},
  volume = {89},
  pages = {033009},
  numpages = {28},
  year = {2014},
  month = 2,
  publisher = {American Physical Society},
  doi = {10.1103/PhysRevD.89.033009},
  eprint         = "1302.6246",
  archivePrefix  = "arXiv",
  primaryClass   = "hep-ph",
  reportNumber   = "SMU-HEP-12-23",
}

@article{Ball:2012cx,
      author         = "{NNPDF Collaboration} and Ball, Richard D. and others",
      title          = "{Parton distributions with LHC data}",
      journal        = "Nucl. Phys. B",
      volume         = "867",
      year           = "2013",
      pages          = "244-289",
      doi            = "10.1016/j.nuclphysb.2012.10.003",
      eprint         = "1207.1303",
      archivePrefix  = "arXiv",
      primaryClass   = "hep-ph",
      reportNumber   = "EDINBURGH-2012-08, IFUM-FT-997, FR-PHENO-2012-014,
                        RWTH-TTK-12-25, CERN-PH-TH-2012-037, SFB-CPP-12-47",
      SLACcitation   = "%%CITATION = ARXIV:1207.1303;%%"
}

@article{Kidonakis:2010tc,
      author         = "Kidonakis, Nikolaos",
      title          = "Next-to-next-to-leading logarithm resummation for $s$-channel single top quark
                        production",
      journal        = "Phys. Rev. D",
      volume         = "81",
      pages          = "054028",
      doi            = "10.1103/PhysRevD.81.054028",
      year           = "2010",
      eprint         = "1001.5034",
      archivePrefix  = "arXiv",
      primaryClass   = "hep-ph",
}

@article{Kidonakis:2011wy,
      author         = "Kidonakis, Nikolaos",
      title          = "Next-to-next-to-leading-order collinear and soft gluon
                        corrections for $t$-channel single top quark production",
      journal        = "Phys. Rev. D",
      volume         = "83",
      pages          = "091503",
      doi            = "10.1103/PhysRevD.83.091503",
      year           = "2011",
      eprint         = "1103.2792",
      archivePrefix  = "arXiv",
      primaryClass   = "hep-ph",
}

@article{Alioli:2009je,
      author         = "Alioli, Simone and Nason, Paolo and Oleari, Carlo and Re, Emanuele",
      title          = "{NLO single-top production matched with shower in POWHEG:
                        \(s\)- and \(t\)-channel contributions}",
      journal        = "JHEP",
      volume         = "09",
      year           = "2009",
      pages          = "111",
      doi            = "10.1088/1126-6708/2009/09/111",
      eprint         = "0907.4076",
      archivePrefix  = "arXiv",
      primaryClass   = "hep-ph",
      SLACcitation   = "%%CITATION = ARXIV:0907.4076;%%",
      related        = "Alioli:2009je-err",
      relatedstring  = "Erratum:",
}

@article{Frederix:2012dh,
      author         = "Frederix, Rikkert and Re, Emanuele and Torrielli, Paolo",
      title          = "{Single-top \(t\)-channel hadroproduction in the four-flavour
                        scheme with POWHEG and aMC@NLO}",
      journal        = "JHEP",
      volume         = "09",
      year           = "2012",
      pages          = "130",
      doi            = "10.1007/JHEP09(2012)130",
      eprint         = "1207.5391",
      archivePrefix  = "arXiv",
      primaryClass   = "hep-ph",
      reportNumber   = "ZU-TH-14-12, IPPP-12-54, DCPT-12-108,
                        CERN-PH-TH-2012-206, MCNET-12-10, LPN12-084",
      SLACcitation   = "%%CITATION = ARXIV:1207.5391;%%"
}

@article{Gleisberg:2008ta,
      author         = "Gleisberg, T. and others",
      title          = "{Event generation with SHERPA 1.1}",
      journal        = "JHEP",
      volume         = "02",
      pages          = "007",
      doi            = "10.1088/1126-6708/2009/02/007",
      year           = "2009",
      eprint         = "0811.4622",
      archivePrefix  = "arXiv",
      primaryClass   = "hep-ph",
}

@article{Hoeche:2009rj,
      author         = "H{\"o}che, Stefan and Krauss, Frank and Schumann, Steffen
                        and Siegert, Frank",
      title          = "{QCD matrix elements and truncated showers}",
      journal        = "JHEP",
      volume         = "05",
      pages          = "053",
      doi            = "10.1088/1126-6708/2009/05/053",
      year           = "2009",
      eprint         = "0903.1219",
      archivePrefix  = "arXiv",
      primaryClass   = "hep-ph",
}

@article{Schumann:2007mg,
      author         = "Schumann, Steffen and Krauss, Frank",
      title          = "{A parton shower algorithm based on Catani--Seymour dipole
                        factorisation}",
      journal        = "JHEP",
      volume         = "03",
      pages          = "038",
      doi            = "10.1088/1126-6708/2008/03/038",
      year           = "2008",
      eprint         = "0709.1027",
      archivePrefix  = "arXiv",
      primaryClass   = "hep-ph",
}

@article{Mangano:2002ea,
      author         = "Mangano, Michelangelo L. and Moretti, Mauro and
                        Piccinini, Fulvio and Pittau, Roberto and Polosa, Antonio
                        D.",
      title          = "{ALPGEN, a generator for hard multiparton processes in
                        hadronic collisions}",
      journal        = "JHEP",
      volume         = "07",
      year           = "2003",
      pages          = "001",
      doi            = "10.1088/1126-6708/2003/07/001",
      eprint         = "hep-ph/0206293",
      archivePrefix  = "arXiv",
      primaryClass   = "hep-ph",
      reportNumber   = "CERN-TH-2002-129, FTN-T-2002-06",
}

@Article{Caciari:2008gp,
     author    = "Cacciari, Matteo and Salam, Gavin P. and Soyez, Gregory",
     title     = "{The anti-$k_{t}$ jet clustering algorithm}",
     journal   = "JHEP",
     volume    = "04",
     year      = "2008",
     pages     = "063",
     eprint    = "0802.1189",
     archivePrefix = "arXiv",
     primaryClass  =  "hep-ph",
     doi       = "10.1088/1126-6708/2008/04/063",
}

@Article{Fastjet,
      author         = "Cacciari, Matteo and Salam, Gavin P. and Soyez, Gregory",
      title          = "{FastJet user manual}",
      journal        = "Eur. Phys. J. C",
      volume         = "72",
      year           = "2012",
      pages          = "1896",
      doi            = "10.1140/epjc/s10052-012-1896-2",
      eprint         = "1111.6097",
      archivePrefix  = "arXiv",
      primaryClass   = "hep-ph",
      reportNumber   = "CERN-PH-TH-2011-297",
      SLACcitation   = "%%CITATION = ARXIV:1111.6097;%%"
}

@article{Feindt:2006pm,
      author         = "Feindt, M. and Kerzel, U.",
      title          = "{The NeuroBayes neural network package}",
      journal        = "Nucl. Instrum. Meth. A",
      volume         = "559",
      pages          = "190-194",
      doi            = "10.1016/j.nima.2005.11.166",
      year           = "2006",
}

@article{Feindt:2004,
      author         = "Feindt, M.",
      title          = "{A Neural Bayesian Estimator for Conditional Probability Densities}",
      year           = "2004",
      eprint         = "physics/0402093",
      archivePrefix  = "arXiv",
      primaryClass   = "physics.data-an",
}

@article{Barlow:1993dm,
      author         = "Barlow, Roger J. and Beeston, Christine",
      title          = "{Fitting using finite Monte Carlo samples}",
      journal        = "Comput. Phys. Commun.",
      volume         = "77",
      pages          = "219-228",
      doi            = "10.1016/0010-4655(93)90005-W",
      year           = "1993",
      reportNumber   = "MAN-HEP-93-1",
}

@Booklet{Conway:2011in,
      author         = "Conway, J.S.",
      title          = "{Incorporating Nuisance Parameters in Likelihoods for
                        Multisource Spectra}",
      year           = "2011",
      eprint         = "1103.0354",
      archivePrefix  = "arXiv",
      primaryClass   = "physics.data-an",
}

@article{CERN-OPEN-2012-016,
      author         = "Cranmer, Kyle and Lewis, George and Moneta, Lorenzo and
                        Shibata, Akira and Verkerke, Wouter",
      title          = "{HistFactory: A tool for creating statistical models for
                        use with RooFit and RooStats}",
      year           = "2012",
      reportNumber   = "CERN-OPEN-2012-016",
      url            = "https://cds.cern.ch/record/1456844",
}

@Booklet{Moneta:2010pm,
      author         = "Moneta, Lorenzo and Belasco, Kevin and Cranmer, Kyle S.
                        and Kreiss, S. and Lazzaro, Alfio and others",
      title          = "{The \textsc{RooStats} Project}",
      year           = "2010",
      eprint         = "1009.1003",
      archivePrefix  = "arXiv",
      primaryClass   = "physics.data-an",
}

@article{James:1975dr,
      author         = "James, F. and Roos, M.",
      title          = "{Minuit -- a system for function minimization and analysis
                        of the parameter errors and correlations}",
      journal        = "Comput. Phys. Commun.",
      volume         = "10",
      year           = "1975",
      pages          = "343-367",
      doi            = "10.1016/0010-4655(75)90039-9",
      reportNumber   = "CERN-DD-75-20",
}

@article{Cowan:2010js,
      author         = "Cowan, Glen and Cranmer, Kyle and Gross, Eilam and
                        Vitells, Ofer",
      title          = "{Asymptotic formulae for likelihood-based tests of new
                        physics}",
      journal        = "Eur. Phys. J. C",
      volume         = "71",
      year           = "2011",
      pages          = "1554",
      doi            = "10.1140/epjc/s10052-011-1554-0",
      eprint         = "1007.1727",
      archivePrefix  = "arXiv",
      primaryClass   = "physics.data-an",
      SLACcitation   = "%%CITATION = ARXIV:1007.1727;%%",
      related        = "Cowan:2010js-err",
      relatedstring  = "Erratum:",
}

@Article{PERF-2007-01,
    author         = "{ATLAS Collaboration}",
    title          = "{The ATLAS Experiment at the CERN Large Hadron Collider}",
    journal        = "JINST",
    volume         = "3",
    year           = "2008",
    pages          = "S08003",
    doi            = "10.1088/1748-0221/3/08/S08003",
    primaryClass   = "hep-ex",
}

@Article{PERF-2010-04,
    author         = "{ATLAS Collaboration}",
    title          = "{Electron performance measurements with the ATLAS detector using the 2010 LHC proton--proton collision data}",
    journal        = "Eur. Phys. J. C",
    volume         = "72",
    year           = "2012",
    pages          = "1909",
    doi            = "10.1140/epjc/s10052-012-1909-1",
    reportNumber   = "CERN-PH-EP-2011-117",
    eprint         = "1110.3174",
    archivePrefix  = "arXiv",
    primaryClass   = "hep-ex",
}

@Article{SOFT-2010-01,
    author         = "{ATLAS Collaboration}",
    title          = "{The ATLAS Simulation Infrastructure}",
    journal        = "Eur. Phys. J. C",
    volume         = "70",
    year           = "2010",
    pages          = "823",
    doi            = "10.1140/epjc/s10052-010-1429-9",
    eprint         = "1005.4568",
    archivePrefix  = "arXiv",
    primaryClass   = "physics.ins-det",
}

@Article{PERF-2011-02,
    author         = "{ATLAS Collaboration}",
    title          = "{Performance of the ATLAS Trigger System in 2010}",
    journal        = "Eur. Phys. J. C",
    volume         = "72",
    year           = "2012",
    pages          = "1849",
    doi            = "10.1140/epjc/s10052-011-1849-1",
    reportNumber   = "CERN-PH-EP-2011-078",
    eprint         = "1110.1530",
    archivePrefix  = "arXiv",
    primaryClass   = "hep-ex",
}

@Article{TOPQ-2011-17,
    author         = "{ATLAS Collaboration}",
    title          = "{Evidence for the associated production of a \(W\) boson and a top quark in ATLAS at \(\sqrt{s} = 7\,\text{TeV}\)}",
    journal        = "Phys. Lett. B",
    volume         = "716",
    year           = "2012",
    pages          = "142",
    doi            = "10.1016/j.physletb.2012.08.011",
    reportNumber   = "CERN-PH-EP-2012-117",
    eprint         = "1205.5764",
    archivePrefix  = "arXiv",
    primaryClass   = "hep-ex",
}

@Article{PERF-2012-01,
    author         = "{ATLAS Collaboration}",
    title          = "{Jet energy measurement and its systematic uncertainty in proton--proton collisions at \(\sqrt{s} = 7\,\text{TeV}\) with the ATLAS detector}",
    journal        = "Eur. Phys. J. C",
    volume         = "75",
    year           = "2015",
    pages          = "17",
    doi            = "10.1140/epjc/s10052-014-3190-y",
    reportNumber   = "CERN-PH-EP-2013-222",
    eprint         = "1406.0076",
    archivePrefix  = "arXiv",
    primaryClass   = "hep-ex",
}

@Article{PERF-2012-04,
    author         = "{ATLAS Collaboration}",
    title          = "{Performance of \(b\)-jet identification in the ATLAS experiment}",
    journal        = "JINST",
    volume         = "11",
    year           = "2016",
    pages          = "P04008",
    doi            = "10.1088/1748-0221/11/04/P04008",
    reportNumber   = "CERN-PH-EP-2015-216",
    eprint         = "1512.01094",
    archivePrefix  = "arXiv",
    primaryClass   = "hep-ex",
}

@Article{TOPQ-2012-20,
    author         = "{ATLAS Collaboration}",
    title          = "{Measurement of the production cross-section of a single top quark in association with a \(W\) boson at \(8\,\text{TeV}\) with the ATLAS experiment}",
    journal        = "JHEP",
    volume         = "01",
    year           = "2016",
    pages          = "064",
    doi            = "10.1007/JHEP01(2016)064",
    eprint         = "1510.03752",
    archivePrefix  = "arXiv",
    primaryClass   = "hep-ex",
}

@Article{TOPQ-2012-21,
    author         = "{ATLAS Collaboration}",
    title          = "{Comprehensive measurements of \(t\)-channel single top-quark production cross sections at \(\sqrt{s} = 7\,\text{TeV}\) with the ATLAS detector}",
    journal        = "Phys. Rev. D",
    volume         = "90",
    year           = "2014",
    pages          = "112006",
    doi            = "10.1103/PhysRevD.90.112006",
    reportNumber   = "CERN-PH-EP-2014-133",
    eprint         = "1406.7844",
    archivePrefix  = "arXiv",
    primaryClass   = "hep-ex",
}

@Article{TRIG-2012-03,
    author         = "{ATLAS Collaboration}",
    title          = "{Performance of the ATLAS muon trigger in \(pp\) collisions at \(\sqrt{s} = 8\,\text{TeV}\)}",
    journal        = "Eur. Phys. J. C",
    volume         = "75",
    year           = "2015",
    pages          = "120",
    doi            = "10.1140/epjc/s10052-015-3325-9",
    reportNumber   = "CERN-PH-EP-2014-154",
    eprint         = "1408.3179",
    archivePrefix  = "arXiv",
    primaryClass   = "hep-ex",
}

@Article{DAPR-2013-01,
    author         = "{ATLAS Collaboration}",
    title          = "{Luminosity determination in \(pp\) collisions at \(\sqrt{s} = 8\,\text{TeV}\) using the ATLAS detector at the LHC}",
    journal        = "Eur. Phys. J. C",
    volume         = "76",
    year           = "2016",
    pages          = "653",
    doi            = "10.1140/epjc/s10052-016-4466-1",
    eprint         = "1608.03953",
    archivePrefix  = "arXiv",
    primaryClass   = "hep-ex",
}

@Article{PERF-2013-03,
    author         = "{ATLAS Collaboration}",
    title          = "{Electron reconstruction and identification efficiency measurements with the ATLAS detector using the 2011 LHC proton--proton collision data}",
    journal        = "Eur. Phys. J. C",
    volume         = "74",
    year           = "2014",
    pages          = "2941",
    doi            = "10.1140/epjc/s10052-014-2941-0",
    reportNumber   = "CERN-PH-EP-2014-040",
    eprint         = "1404.2240",
    archivePrefix  = "arXiv",
    primaryClass   = "hep-ex",
}

@Article{PERF-2014-02,
    author         = "{ATLAS Collaboration}",
    title          = "{Determination of jet calibration and energy resolution in proton--proton collisions at \(\sqrt{s} = 8\,\text{TeV}\) using the ATLAS detector}",
    year           = "2019",
    reportNumber   = "CERN-EP-2019-057",
    eprint         = "1910.04482",
    archivePrefix  = "arXiv",
    primaryClass   = "hep-ex",
}

@Article{PERF-2014-03,
    author         = "{ATLAS Collaboration}",
    title          = "{Performance of pile-up mitigation techniques for jets in \(pp\) collisions at \(\sqrt{s} = 8\,\text{TeV}\) using the ATLAS detector}",
    journal        = "Eur. Phys. J. C",
    volume         = "76",
    year           = "2016",
    pages          = "581",
    doi            = "10.1140/epjc/s10052-016-4395-z",
    reportNumber   = "CERN-PH-EP-2015-206",
    eprint         = "1510.03823",
    archivePrefix  = "arXiv",
    primaryClass   = "hep-ex",
}

@Article{PERF-2014-04,
    author         = "{ATLAS Collaboration}",
    title          = "{Performance of algorithms that reconstruct missing transverse momentum in \(\sqrt{s} = 8\,\text{TeV}\) proton--proton collisions in the ATLAS detector}",
    journal        = "Eur. Phys. J. C",
    volume         = "77",
    year           = "2017",
    pages          = "241",
    doi            = "10.1140/epjc/s10052-017-4780-2",
    reportNumber   = "CERN-EP-2016-134",
    eprint         = "1609.09324",
    archivePrefix  = "arXiv",
    primaryClass   = "hep-ex",
}

@Article{PERF-2014-05,
    author         = "{ATLAS Collaboration}",
    title          = "{Measurement of the muon reconstruction performance of the ATLAS detector using 2011 and 2012 LHC proton--proton collision data}",
    journal        = "Eur. Phys. J. C",
    volume         = "74",
    year           = "2014",
    pages          = "3130",
    doi            = "10.1140/epjc/s10052-014-3130-x",
    reportNumber   = "CERN-PH-EP-2014-151",
    eprint         = "1407.3935",
    archivePrefix  = "arXiv",
    primaryClass   = "hep-ex",
}

@Article{PERF-2014-07,
    author         = "{ATLAS Collaboration}",
    title          = "{Topological cell clustering in the ATLAS calorimeters and its performance in LHC Run~1}",
    journal        = "Eur. Phys. J. C",
    volume         = "77",
    year           = "2017",
    pages          = "490",
    doi            = "10.1140/epjc/s10052-017-5004-5",
    reportNumber   = "CERN-PH-EP-2015-304",
    eprint         = "1603.02934",
    archivePrefix  = "arXiv",
    primaryClass   = "hep-ex",
}

@Article{PERF-2015-01,
    author         = "{ATLAS Collaboration}",
    title          = "{Reconstruction of primary vertices at the ATLAS experiment in Run~1 proton--proton collisions at the LHC}",
    journal        = "Eur. Phys. J. C",
    volume         = "77",
    year           = "2017",
    pages          = "332",
    doi            = "10.1140/epjc/s10052-017-4887-5",
    reportNumber   = "CERN-EP-2016-150",
    eprint         = "1611.10235",
    archivePrefix  = "arXiv",
    primaryClass   = "hep-ex",
}

@Article{TOPQ-2015-16,
    author         = "{ATLAS Collaboration}",
    title          = "{Measurement of the cross-section for producing a \(W\) boson in association with a single top quark in \(pp\) collisions at \(\sqrt{s} = 13\,\text{TeV}\) with ATLAS}",
    journal        = "JHEP",
    volume         = "01",
    year           = "2018",
    pages          = "063",
    doi            = "10.1007/JHEP01(2018)063",
    reportNumber   = "CERN-EP-2016-238",
    eprint         = "1612.07231",
    archivePrefix  = "arXiv",
    primaryClass   = "hep-ex",
}

@Article{PERF-2016-01,
    author         = "{ATLAS Collaboration}",
    title          = "{Electron efficiency measurements with the ATLAS detector using 2012 LHC proton--proton collision data}",
    journal        = "Eur. Phys. J. C",
    volume         = "77",
    year           = "2017",
    pages          = "195",
    doi            = "10.1140/epjc/s10052-017-4756-2",
    reportNumber   = "CERN-EP-2016-262",
    eprint         = "1612.01456",
    archivePrefix  = "arXiv",
    primaryClass   = "hep-ex",
}

@Article{SUSY-2016-16,
    author         = "{ATLAS Collaboration}",
    title          = "{Search for top-squark pair production in final states with one lepton, jets, and missing transverse momentum using \(36\,\text{fb}^{-1}\) of \(\sqrt{s} = 13\,\text{TeV}\) \(pp\) collision data with the ATLAS detector}",
    journal        = "JHEP",
    volume         = "06",
    year           = "2018",
    pages          = "108",
    doi            = "10.1007/JHEP06(2018)108",
    reportNumber   = "CERN-EP-2017-246",
    eprint         = "1711.11520",
    archivePrefix  = "arXiv",
    primaryClass   = "hep-ex",
}

@Article{SUSY-2016-29,
    author         = "{ATLAS Collaboration}",
    title          = "{Search for \(B-L\) \(R\)-parity-violating top squarks in \(\sqrt{s} = 13\,\text{TeV}\) \(pp\) collisions with the ATLAS experiment}",
    journal        = "Phys. Rev. D",
    volume         = "97",
    year           = "2018",
    pages          = "032003",
    doi            = "10.1103/PhysRevD.97.032003",
    reportNumber   = "CERN-EP-2017-171",
    eprint         = "1710.05544",
    archivePrefix  = "arXiv",
    primaryClass   = "hep-ex",
}

@Article{TOPQ-2016-12,
    author         = "{ATLAS Collaboration}",
    title          = "{Measurement of differential cross-sections of a single top quark produced in association with a \(W\) boson at \(\sqrt{s} = 13\,\text{TeV}\) with ATLAS}",
    journal        = "Eur. Phys. J. C",
    volume         = "78",
    year           = "2018",
    pages          = "186",
    doi            = "10.1140/epjc/s10052-018-5649-8",
    reportNumber   = "CERN-EP-2017-221",
    eprint         = "1712.01602",
    archivePrefix  = "arXiv",
    primaryClass   = "hep-ex",
}

@Article{TOPQ-2017-05,
    author         = "{ATLAS Collaboration}",
    title          = "{Probing the Quantum Interference between Singly and Doubly Resonant Top-Quark Production in \(pp\) Collisions at \(\sqrt{s} = 13\,\text{TeV}\) with the ATLAS Detector}",
    journal        = "Phys. Rev. Lett.",
    volume         = "121",
    year           = "2018",
    pages          = "152002",
    doi            = "10.1103/PhysRevLett.121.152002",
    reportNumber   = "CERN-EP-2018-087",
    eprint         = "1806.04667",
    archivePrefix  = "arXiv",
    primaryClass   = "hep-ex",
}

@Booklet{ATL-SOFT-PUB-2020-001,
    author         = "{ATLAS Collaboration}",
    title          = "{ATLAS Computing Acknowledgements}",
    howpublished   = "{ATL-SOFT-PUB-2020-001}",
    url            = "https://cds.cern.ch/record/2717821",
}

@Article{CMS-TOP-11-022,
    author         = "{CMS Collaboration}",
    title          = "{Evidence for associated production of a single top quark and \(W\) boson in \(pp\) collisions at \(\sqrt{s} = 7\,\text{TeV}\)}",
    journal        = "Phys. Rev. Lett.",
    volume         = "110",
    year           = "2013",
    pages          = "022003",
    doi            = "10.1103/PhysRevLett.110.022003",
    reportNumber   = "CERN-PH-EP-2012-266",
    eprint         = "1209.3489",
    archivePrefix  = "arXiv",
    primaryClass   = "hep-ex",
}

@Article{CMS-TOP-12-040,
    author         = "{CMS Collaboration}",
    title          = "{Observation of the Associated Production of a Single Top Quark and a \(W\) Boson in \(pp\) Collisions at \(\sqrt{s} = 8\,\text{TeV}\)}",
    journal        = "Phys. Rev. Lett.",
    volume         = "112",
    year           = "2014",
    pages          = "231802",
    doi            = "10.1103/PhysRevLett.112.231802",
    reportNumber   = "CERN-PH-EP-2013-237",
    eprint         = "1401.2942",
    archivePrefix  = "arXiv",
    primaryClass   = "hep-ex",
}

@Article{CMS-TOP-17-018,
    author         = "{CMS Collaboration}",
    title          = "{Measurement of the production cross section for single top quarks in association with \(W\) bosons in proton--proton collisions at \(\sqrt{s} = 13\,\text{TeV}\)}",
    journal        = "JHEP",
    volume         = "10",
    year           = "2018",
    pages          = "117",
    doi            = "10.1007/JHEP10(2018)117",
    reportNumber   = "CERN-EP-2018-074",
    eprint         = "1805.07399",
    archivePrefix  = "arXiv",
    primaryClass   = "hep-ex",
}

@Booklet{ATLAS-CONF-2014-058,
    author         = "{ATLAS Collaboration}",
    title          = "{Estimation of non-prompt and fake lepton backgrounds in final states with top quarks produced in proton--proton collisions at \(\sqrt{s} = 8~\text{TeV}\) with the ATLAS Detector}",
    howpublished   = "{ATLAS-CONF-2014-058}",
    url            = "https://cds.cern.ch/record/1951336",
    year           = "2014",
}

@Booklet{ATL-PHYS-PUB-2011-008,
    author         = "{ATLAS Collaboration}",
    title          = "{New ATLAS event generator tunes to 2010 data}",
    howpublished   = "{ATL-PHYS-PUB-2011-008}",
    url            = "https://cds.cern.ch/record/1345343",
    year           = "2011",
}

@Booklet{ATL-PHYS-PUB-2015-010,
    author         = "{ATLAS Collaboration}",
    title          = "{Study of correlation of PDF uncertainty in single top and top pair production at the LHC}",
    howpublished   = "{ATL-PHYS-PUB-2015-010}",
    url            = "https://cds.cern.ch/record/2020601",
    year           = "2015",
}
 
\clearpage
 
\begin{flushleft}
\hypersetup{urlcolor=black}
{\Large The ATLAS Collaboration}

\bigskip

\AtlasOrcid[0000-0002-6665-4934]{G.~Aad}$^\textrm{\scriptsize 102}$,    
\AtlasOrcid[0000-0002-5888-2734]{B.~Abbott}$^\textrm{\scriptsize 130}$,    
\AtlasOrcid[0000-0002-7248-3203]{D.C.~Abbott}$^\textrm{\scriptsize 103}$,    
\AtlasOrcid[0000-0002-2788-3822]{A.~Abed~Abud}$^\textrm{\scriptsize 71a,71b}$,    
\AtlasOrcid[0000-0002-1002-1652]{K.~Abeling}$^\textrm{\scriptsize 53}$,    
\AtlasOrcid[0000-0002-2987-4006]{D.K.~Abhayasinghe}$^\textrm{\scriptsize 94}$,    
\AtlasOrcid[0000-0002-8496-9294]{S.H.~Abidi}$^\textrm{\scriptsize 169}$,    
\AtlasOrcid[0000-0002-8279-9324]{O.S.~AbouZeid}$^\textrm{\scriptsize 40}$,    
\AtlasOrcid{N.L.~Abraham}$^\textrm{\scriptsize 158}$,    
\AtlasOrcid[0000-0001-5329-6640]{H.~Abramowicz}$^\textrm{\scriptsize 163}$
% ...
\bigskip
\\

% ...
$^{40}$Niels Bohr Institute, University of Copenhagen, Copenhagen; Denmark.\\
% ...
$^{53}$II. Physikalisches Institut, Georg-August-Universit\"{a}t G\"ottingen, G\"ottingen; Germany.\\
% ...
$^{71}$$^{(a)}$INFN Sezione di Pavia;$^{(b)}$Dipartimento di Fisica, Universit\`a di Pavia, Pavia; Italy.\\
% ...
$^{94}$Department of Physics, Royal Holloway University of London, Egham; United Kingdom.\\
% ...
$^{102}$CPPM, Aix-Marseille Universit\'e, CNRS/IN2P3, Marseille; France.\\
$^{103}$Department of Physics, University of Massachusetts, Amherst MA; United States of America.\\
% ...
$^{130}$Homer L. Dodge Department of Physics and Astronomy, University of Oklahoma, Norman OK; United States of America.\\
% ...
$^{158}$Department of Physics and Astronomy, University of Sussex, Brighton; United Kingdom.\\
% ...
$^{163}$Raymond and Beverly Sackler School of Physics and Astronomy, Tel Aviv University, Tel Aviv; Israel.\\
% ...
$^{169}$Department of Physics, University of Toronto, Toronto ON; Canada.\\
% ...
$^{a}$ Also at Borough of Manhattan Community College, City University of New York, New York NY; United States of America.\\
$^{b}$ Also at CERN, Geneva; Switzerland.\\
% ...

\end{flushleft}


\end{document}